\documentclass[namedreferences]{solarphysics}
\usepackage[utf8x]{inputenc}
\usepackage[hyperref,optionalrh,showbiblabels]{spr-sola-addons} % For Solar Physics 
\usepackage{graphicx}        % For eps figures, newer & more powerfull
\usepackage{color}           % For color text: \color command
\usepackage{breakurl}        % For breaking URLs easily trough lines
\def\dpar#1#2{\frac{\partial#1}{\partial#2}}
\renewcommand{\vec}[1]{{\mathbfit #1}}

\newcommand{\curl}{ {\bf \nabla} \times}

\newcommand{\bb}{\vec B}

\begin{document}

\begin{article}
%\tracingmacros=2
\begin{opening}

\title{{Three-Dimensional Reconstruction and Thermal Modelling of Observed Loops}\\}

\author[addressref={aff1,aff2},corref,email={federico@iafe.uba.ar}]{\inits{F.A.}\fnm{Federico A.}~\lnm{Nuevo}}%\sep
\author[addressref={aff1,aff3},email={cmaccormack@iafe.uba.ar}]{\inits{C.}\fnm{Cecilia}~\lnm{Mac Cormack}}%\sep
\author[addressref={aff1},email={lopezf@iafe.uba.ar}]{\inits{M.C.}\fnm{Marcelo C.}~\lnm{L\'opez Fuentes}}%\sep
\author[addressref={aff1,aff3},email={albert@iafe.uba.ar}]{\inits{A.M.}\fnm{Alberto M.}~\lnm{V\'asquez}}%\sep
\author[addressref={aff1,aff4},email={mandrini@iafe.uba.ar}]{\inits{C.H.}\fnm{Cristina H.}~\lnm{Mandrini}}%\sep

\address[id=aff1]{Instituto de Astronom\'ia y F\'isica del Espacio (IAFE) CONICET-UBA, CC 67 - Suc 28, (C1428ZAA) Ciudad Aut\'onoma de Buenos Aires, Argentina.}

\address[id=aff2]{Ciclo B\'asico Com\'un (CBC), Universidad de Buenos Aires (UBA), Ciudad Aut\'onoma de Buenos Aires, Argentina.}

\address[id=aff3]{Universidad Nacional de Tres de Febrero (UNTREF). Departamento de Ciencia y Tecnolog\'ia, S\'aenz Pe\~na, Argentina.}

\address[id=aff4]{Departamento de F\'isica, Facultad de Ciencias Exactas y Naturales (FCEN), Universidad de Buenos Aires (UBA), Pabell\'on I, Ciudad Universitaria (C1428ZAA) Ciudad Aut\'onoma de Buenos Aires, Argentina.}

\runningauthor{F.A. Nuevo et al.}
\runningtitle{3D Resconstruction of Loop Density and Temperature}

\begin{abstract}
Due to their characteristic temperature and density, loop structures in active regions (ARs) can be seen bright in extreme ultraviolet (EUV) and soft X-ray images. The semiempirical determination of the three-dimensional (3D) distribution of basic physical parameters (electronic density and temperature, and magnetic field) is a key constraint for coronal heating models. In this work we develop a technique for the study of EUV bright loops based on differential emission measure (DEM) analysis and we first apply it to  AR structures observed by the {Atmospheric Imaging Assembly} (AIA) on board the {Solar Dynamics Observatory} (SDO). The 3D structure and intensity of the magnetic field of the observed EUV loops are modelled using force-free field extrapolations based on magnetograms taken by the {Helioseismic and Magnetic Imager} (HMI) on board SDO. In this work we report the results obtained for several bright loops identified in different ARs. Our analysis indicates that the mean and width of the temperature distributions are nearly invariant along the loop lengths. For a particular loop we study its temporal evolution and find that these characteristics remain approximately constant for most of its life time. The appearance and disappearance of this loop occurs at time-scales much shorter than its life time of $\approx 2.5$ hours. The results of this analysis are compared with numerical simulations using the zero-dimensional (0D) hydrodynamic model, Enthalpy-Based Thermal Evolution of Loops (EBTEL). We study two alternative heating scenarios: first, we apply a constant heating rate assuming loops in quasi-static equilibrium, and second, we heat the loops using impulsive events or nanoflares. We find that all the observed loops are overdense with respect to a quasi-static equilibrium solution and that the nanoflare heating better reproduces the observed densities and temperatures.
\end{abstract}
\end{opening} 

\tracingmacros=0

%%%%%%%%%%%%%%%%%%%%%%%%%INTRODUCTION%%%%%%%%%%%%%%%%%%%%%%%%%%%%%%%%%5

\section{Introduction}
     \label{S-Introduction} 

The solar corona is a highly inhomogeneous plasma structured by the magnetic field. It is formed by magnetic flux tubes where the plasma is confined. These building blocks of the solar corona are called loops. In active regions (ARs), where the magnetic field is very intense, some loops are visible in extreme ultraviolet (EUV) and soft X-ray images, due to their characteristic temperature and density.
     
AR loops are usually classified as warm and hot \citep{Reale2014}. While the hot loops, observed in soft X-ray images, are located in the AR core and have temperatures above 2 MK, warm loops are observed in the EUV range, are located in the AR periphery, and have temperatures between 1.0 to 1.5 MK. A third category is that of cool loops below 1.0 MK.

The first studies of AR loops using soft X-ray observations from {Skylab} showed that hot loops are consistent with static or quasi-static equilibrium \citep{Rosner1978}.
Later studies based on high spatial and temporal resolution observations in the EUV range, as those provided by the {Transition Region and Coronal Explorer} (TRACE) showed that warm loops are {typically} not consistent with quasi-static equilibrium. Their characteristic density and scale height are larger than those predicted by static equilibrium conditions for their respective temperatures \citep{Aschwanden2001,Winebarger2003}. 

Different mechanisms were proposed to explain overdensity in warm loops. One plausible scenario contemplates quasi-steady heating concentrated at the loop footpoints (near the chromospheric base). This kind of mechanism can reproduce the observed densities and temperatures, but the loop evolves to a thermal non-equilibrium producing important asymmetries in the density distribution along {it} \citep{Lionello2013,Klimchuk2010}. In another scenario the loop is heated by an impulsive mechanism along its coronal part ({e.g.} in the form of nanoflares). This mechanism also reproduces the observed densities and temperatures, but the predicted temporal evolution of the loop results more dynamic than actually observed \citep{klimchuk15}. This discrepancy can be explained if the loop is made of unresolved strands that evolve independently (see {e.g.} \citealt{LopezFuentes2007}).

In this article, we select and analyse a series of active region (AR) loops observed with the {Atmospheric Imaging Assembly} (AIA: \citealt{lemen_2012}) on board the Solar Dynamics Observatory (SDO, \citealt{pesnell_2012}). We develop a differential emission measure (DEM) technique to analyse EUV images and reconstruct the electron temperature and density along the observed loops. We use linear force-free field (LFFF) extrapolations of the AR magnetic field to determine the three-dimensional (3D) structure of the observed loops. As boundary condition for the extrapolations we use magnetograms obtained with the {Helioseismic and Magnetic Imager} (HMI, \citealt{scherrer_2012}) on board SDO. For a particular loop we also study the temporal evolution of the obtained plasma parameters.

{We} study the consistency of the evolution of {the} analysed {loops} with the two scenarios mentioned before: quasi-static equilibrium with a constant heating along {them} and nanoflare heating, we compare our results with the electron density and temperature values predicted by a hydrodynamic (HD) model. To this end, we use the code Enthalpy-Based Thermal Evolution of Loops (EBTEL) \citep{Klimchuk2008}, whose input parameters are the heating rate function and the loop length. 

In Section \ref{method}, we describe the loop selection, DEM analysis, and magnetic field modelling. In Section \ref{res}, we present our results and we study the temporal evolution of one of the analysed loops. In Section \ref{ebtel_sec}, we describe the EBTEL hydrodynamic model and compare our results with the solutions of the model. Finally, we discuss and conclude in Section \ref{S-Summary}.

%%%%%%%%%%%%%%%%%%%%%%METHODOLOGY%%%%%%%%%%%%%%%%%%%%%%%%%%%%%%%%%%%%%%%%

\section{Methodology}
\label{method}

\subsection{EUV Data and Loop Selection}

{Using images provided by the AIA telescope on board the SDO spacecraft, we select as target for this study}
several loops observed at the periphery of various ARs. We choose images in which the target loops are observed at their maximum brightness, determined by inspection of image time-series of each AR.
The selection criterion is that a significant portion, or the totality of the loop, should be visible in the EUV images. To visually select the loops we use the 171 \AA\ AIA channel, corroborating that they are also visible in other channels (e.g. 131 and 193 \AA). 
Table \ref{ARlist}  lists all the loops selected for our study, indicating the AR to which the loop belongs and the date and time of the image where it has been identified. 
\begin{table}
%\begin{center}
\begin{tabular}{cccc}
\hline
 Loop number & AR NOAA &  Date & Time (UT) \\
\hline
 1 & 11130 & 11-29-2010 & 20:56 \\
 2 & 11283 & 09-05-2011 & 20:46 \\
 3 & 11635 & 12-24-2012 & 22:01 \\
 4 & 11652 & 01-10-2013 & 21:55 \\
 5 & 11652 & 01-10-2013 & 21:55 \\
 6 & 11670 & 02-08-2013 & 22:01 \\
 7 & 11670 & 02-08-2013 & 22:01 \\
 8 & 11670 & 02-08-2013 & 22:01 \\
 9 & 12002 & 03-14-2014 & 20:29 \\
 10 & 12291 & 02-23-2015 & 22:30 \\
 11 & 12384 & 07-15-2015 & 02:26 \\
 12 & 12390 & 07-27-2015 & 23:42 \\
 13 & 12390 & 07-27-2015 & 23:42 \\
\hline
\end{tabular}
%\end{center}
\caption{List of magnetic loops selected for this study, indicating the AR to which each loop belongs and the date and time of the AIA 171 image where {it has been} identified.}  
\label{ARlist}
\end{table}

\begin{figure}
      \includegraphics[width=0.44\textwidth]{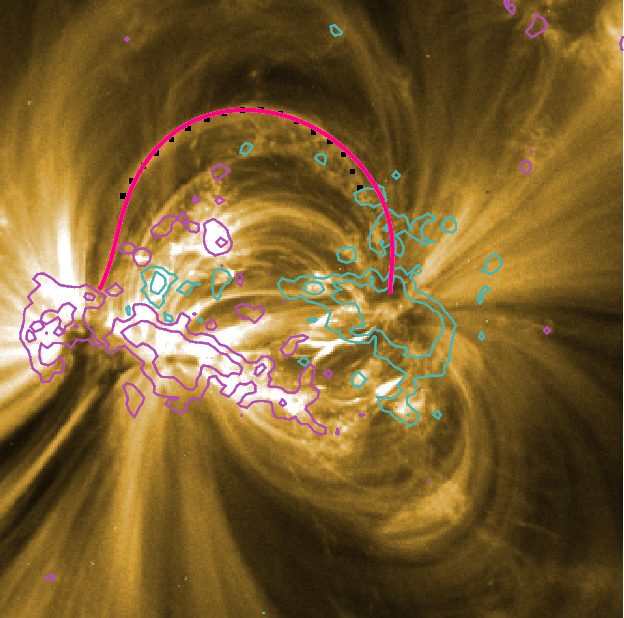}
      \includegraphics[width=0.5\textwidth]{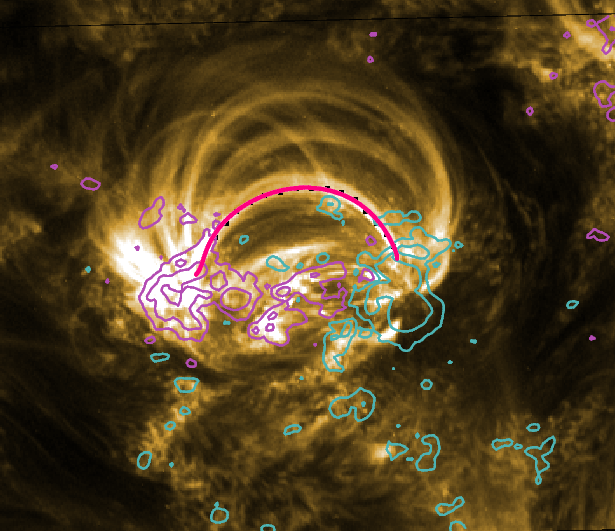}
      \includegraphics[width=0.47\textwidth]{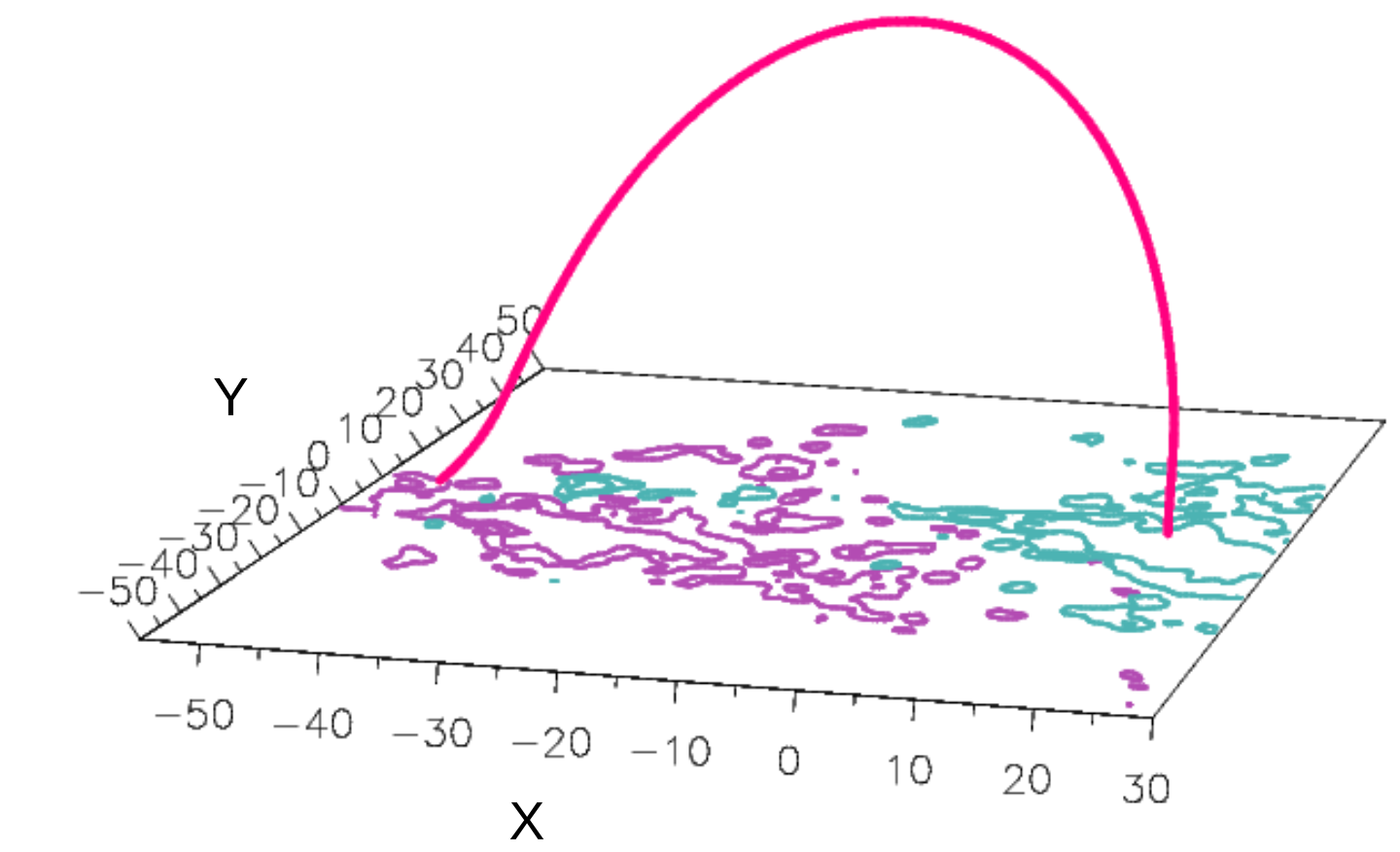}
      \includegraphics[width=0.47\textwidth]{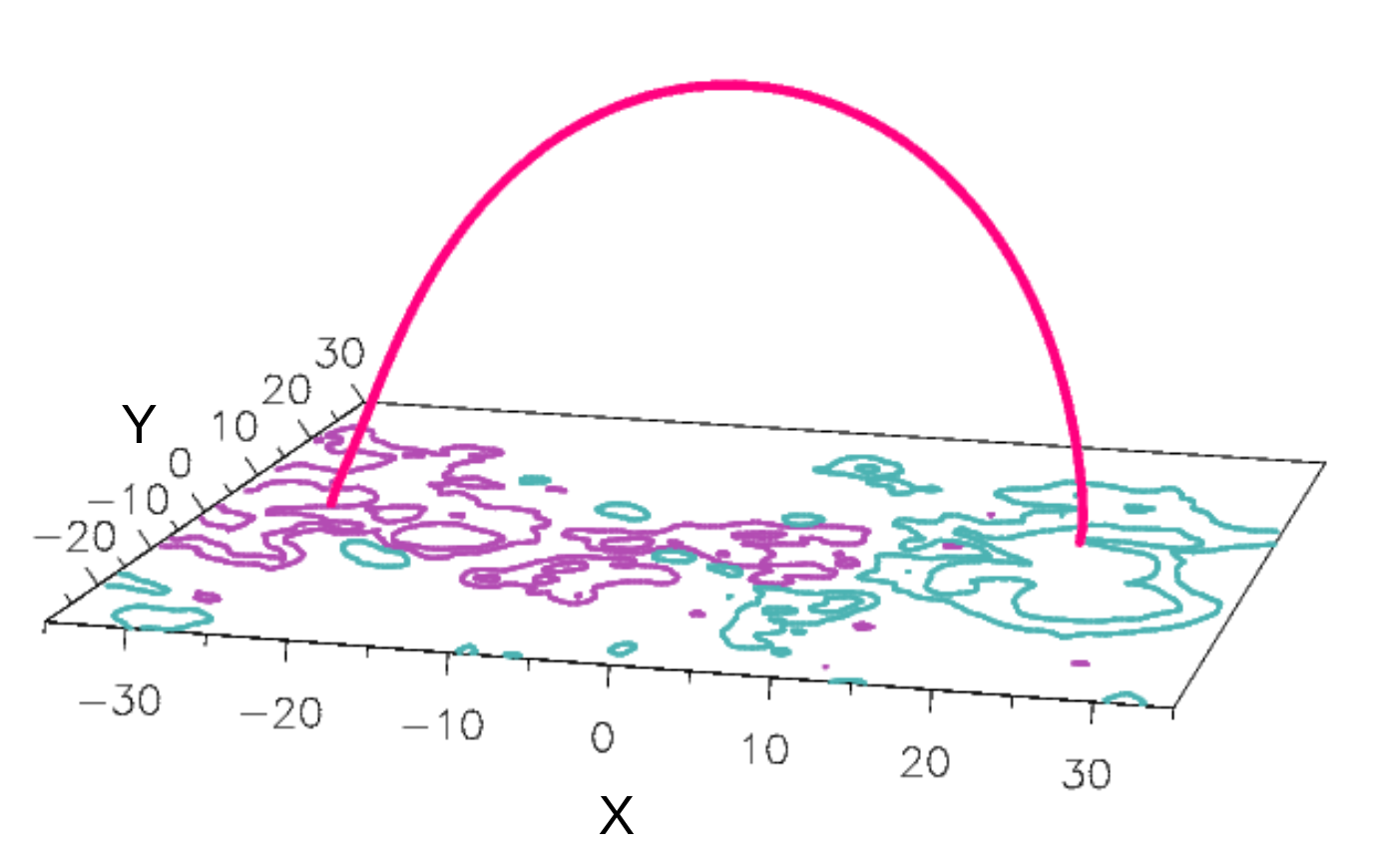}
      \caption{Left panel (top): 171 \AA\ AIA image of AR NOAA 11130 (loop number 1). Magnetic field contours of $\pm$ 100 and $\pm$ 500 G, positive (negative) shown in magenta (cyan) colour, corresponding to the HMI magnetogram are superimposed. The magenta line is the magnetic field line from the LFFF extrapolation that better fits the observed loop (see Section \ref{Bfield_sec}). Right panel (top): Same for loop number 6 observed in AR 11670. Left panel (bottom): The same field line as in the top left panel seen from a different point of view for a better visualisation of its 3D geometry. The value of $\alpha$ for AR 11130 model is $-1.3 \times 10^{-2} {\rm Mm}^{-1}$. Right panel (bottom): Similar to the panel on the left for AR 11670. In this case, the value of $\alpha$ is  $1.3 \times 10^{-2} {\rm Mm}^{-1}$.
      The X and Y axes are the local Cartesian coordinates in Mm in the east to west (E-W) and south to north (S-N) direction, respectively.}
  \label{EUVfig}
\end{figure}

To perform a differential emission measure (DEM) analysis to reconstruct the loop density and temperature, we select locations uniformly distributed along the loop, covering most of its observed length. 
As an example, the left panel of Figure \ref{EUVfig} shows the 171 \AA\ AIA image of AR NOAA 11130 used to identify loop number 1 (see Table \ref{ARlist}). The black dots over the loop are the spatial locations where we perform the DEM analysis. The magenta line corresponds to the field line from the LFFF model that best fits the loop (see Section \ref{Bfield_sec}). As another example, the right panel of Figure \ref{EUVfig} shows the 171 \AA\ AIA image used to identify loop number 6 (AR NOAA 11670) in a similar fashion to the left panel.

%%%%%%%%%%%%%%%%%%%%%%%%%%DEM ANALISIS%%%%%%%%%%%%%%%%%%%%%%%%%%

\subsection{DEM Analysis at Selected Loop Locations}
\label{DEM_sec}

\begin{figure}
      \centering
      \includegraphics[width=\textwidth]{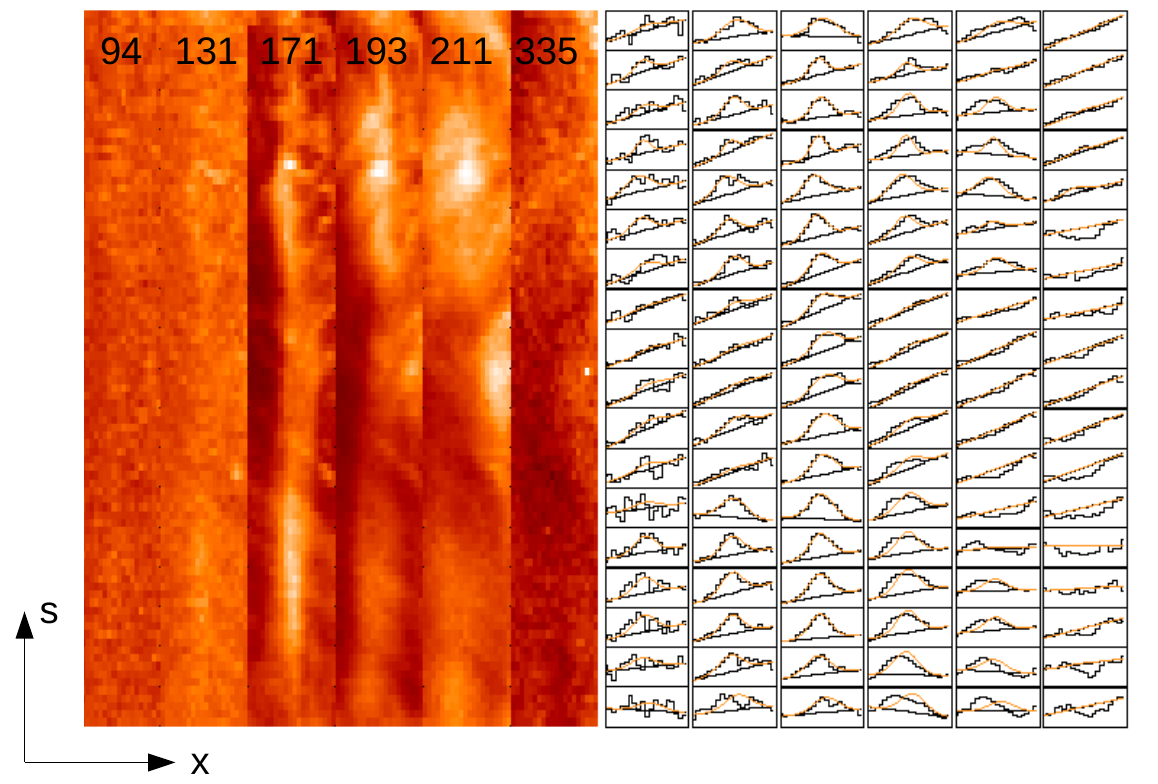}
      \caption{Left panel: Stretched-out version of loop number 1 in the six AIA channels used. From left to right: 94, 131, 171, 193, 211, and 335 \AA. 
      Right panel: Cross-sectional variation of the intensity for each channel (black staircase-like curves) and the corresponding fitted functions given by Equation \ref{gauss+lin} (orange curves). The black staircase-like lines indicate the fit to the background (last two terms in Equation \ref{gauss+lin}). 
      The figure was generated with the open access code {\tt aia\_loop\_autodem.pro} from the SolarSoft software package. {The direction of both the main axis along the loop ($s$) and its perpendicular axis ($x$) in the plane of the sky, which apply to each of the six stretched-out loop images, are indicated as a reference.}}      
   \label{background_loop1}
  \end{figure}

\begin{figure}
      \centering
      \includegraphics[width=\textwidth]{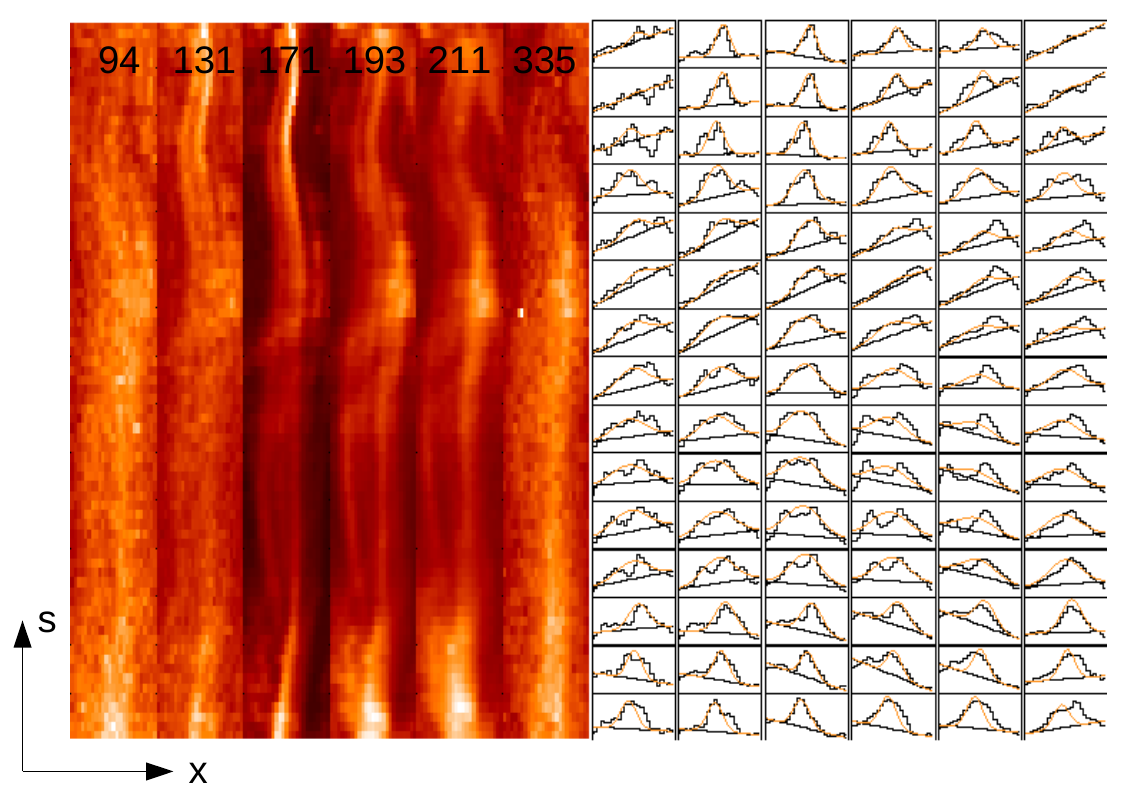}
      \caption{Same as Figure \ref{background_loop1} for loop number 6.}
   \label{background_loop6}
  \end{figure}

To obtain the intrinsic intensity of the loops selected for analysis, we subtract the background intensity following the method developed by \citet{Aschwanden2013}. For each target loop, sample data points (black dots in Figure \ref{EUVfig}) are selected by careful visual inspection. Those dots are then fitted with a polynomial function to determine the main axis of the loop. A bi-linear interpolation scheme is then applied to the image data along and around the main axis of the loop in order to produce a stretched-out image of the loop. This is done for the images provided by the six analysed AIA channels. Examples of stretched-out images for loops 1 and 6 are shown in the left panels of Figures \ref{background_loop1} and \ref{background_loop6}, respectively. As indicated in the images, the main axis of the loop is called $s$, and the perpendicular (cross-sectional) axis in the plane of the sky is called $x$. At each location $s_i$, the intensity of data points $I_{k,i}(x)$ along axis $x$ (with $x$ having units of pixels), for each channel $k$, are measured and fitted to a parametric model of a Gaussian plus a linear function,
\begin{equation}
 I_{k,i}^{\rm mod}(x) = I^{(0)}_{k,i}\; {\rm exp}\left[-\frac{(x-x_{0,i})^2}{2\sigma_{0,i}^2}\right] + I^{(1)}_{k,i}\,x + I^{(2)}_{k,i}\,.
\label{gauss+lin}
\end{equation}
At each point $s_i$, the quantities $I^{(0)}_{k,i}$, $x_{0,i}$, and $\sigma_{0,i}$ are the amplitude, centroid location, and standard deviation of the Gaussian function, respectively. $I^{(1)}_{k,i}$ and $I^{(2)}_{k,i}$ are the slope and intercept of the linear function, respectively. In this model, the Gaussian function represents the steep transverse intensity variation associated to the intrinsic emission of the loop. The linear function represents the smooth transverse intensity variation associated to the background emission of the diffuse corona surrounding the bright loop.

The fitting given by Equation \ref{gauss+lin} is carried out in two steps. The parameters $I^{(1)}_{k,i}$ and $I^{(2)}_{k,i}$ are determined first. The linear function in Equation \ref{gauss+lin} is fit to the data points $I_{k,i}(x)$ along the $x$ axis that are farther from the main axis $s$ and at both sides of it, out of the segment of the axis $x$ that corresponds to the intrinsic emission of the bright loop. The extremes of the segment along $x$ are selected by careful visual inspection point by point along each loop. The background subtracted intensity data points within this segment, $I^{({\rm loop})}_{k,i}(x)$, are then estimated at each pixel $x$ as
\begin{equation}
I^{({\rm loop})}_{k,i}(x) = I_{k,i}(x) - \left( I^{(1)}_{k,i} \, x + I^{(2)}_{k,i} \right). 
\label{intensity_loop}                         % Notar que esto será Eq (2)
\end{equation}
 
In a second step, the Gaussian function of Equation \ref{gauss+lin} is fit to the EUV bright loop intrinsic intensity data points, $I^{({\rm loop})}_{k,i}(x)$, to find the three parameters of the Gaussian function at point $s_i$.

We use the 171 \AA\ channel to identify the EUV loops, as they exhibit strong emission close to $\approx$ 1 MK. The method described above is first applied to this image so that the parameters corresponding to the intrinsic cross-sectional loop intensity for this channel are found at each location $s_i$, namely $I^{(0)}_{171,i}$, $x_{i,0}$ and $\sigma_{i,0}$.
 
For the other channels, the first step of the fitting method is applied to determine the linear function for  the background emission, i.e. to find parameters $I^{(1)}_{k,i}$ and $I^{(2)}_{k,i}$. Equation \ref{intensity_loop} is then used to find the EUV bright loop intrinsic intensity data points $I^{({\rm loop})}_{k,i}(x)$ for each channel. We then estimate the total intrinsic (background subtracted) intensity of the loop in channel $k$ at each location $s_i$ as the integral
\begin{equation}
 S_{k,i} = \int_{x_{0,i}-\sigma_{0,i}}^{x_{0,i}+\sigma_{0,i}} I^{({\rm loop})}_{k,i}(x) \,dx \,.
 \label{Sk_def}
 \end{equation}
Note that the integral is performed over the range of pixels $x$ where the signal-to-noise (S/N) ratio of the EUV intensity is largest.

As the EUV loops are best visible in the 171 \AA\ channel, the centroid and width of the EUV bright loop for all other channels are set to the values found for 171 \AA, i.e. $x_{0,i}$ and $\sigma_{0,i}$, respectively. Finally, the amplitude of the Gaussian model of the EUV loop intrinsic intensity is estimated as
\begin{equation}
 I^{(0)}_{k,i} = I^{(0)}_{171,i}\,\frac{S_{k,i}}{S_{171,i}} \,.
 \label{I_other_bands}
\end{equation}
The motivation for the choice of Equation \ref{I_other_bands}, is that, assuming a coronal filling factor of order one, the amplitude-to-area ratio of the Gaussian model is set uniform for all channels.
  
The right panels of Figure \ref{background_loop1} and \ref{background_loop6} show the cross-sectional intensity profiles (in black colour) and the corresponding Gaussian and linear (drawn as a stair-case like curve) {fits in orange and black colour, respectively, at selected points} along the main axis of loops number 1 and 6.

As a measure of the (dimensionless) fractional uncertainty of the intrinsic intensity $\Delta(I^{(0)}_{k,i})/I^{(0)}_{k,i}$ due to the fit and background subtraction procedure, we compute the difference between the observed and modelled total intrinsic intensity, $S_k$ and $S_k^{\rm mod}$, relative to the former, {i.e.}

\begin{equation}
\Delta(I^{(0)}_{k,i})/I^{(0)}_{k,i} \approx \left( S_{k,i}^{\rm mod} - S_{k,i} \right) / S_{k,i}
\label{back_error}
\end{equation}

\noindent
where the modelled total intrinsic intensity $S_{k,i}^{\rm mod}$ is computed using Equation \ref{Sk_def} but replacing $I_{k,i}(x)$ by $I_{k,i}^{\rm mod}$ in the integrand. In points where the fit is nearly perfect this measure tends to zero. In points where the fit is highly unrepresentative of the observed intensity profile this measure can be as high as the order of one.

After the background emission subtraction method is applied, the intrinsic intensities $I^{(0)}_{k,i}$ for all channels $k$ are used to perform a DEM analysis at each location $s_i$ of each loop.
Applied to image pixels, DEM techniques allow determination of the temperature distribution function, ${\rm DEM}(T)$. This function measures the amount of plasma at temperature $T$ that is along the line-of-sight (LOS) associated to the pixel. Applied to the intrinsic intensities, this DEM function measures the distribution of the emitting plasma located only at the observed bright loop. 
In this work, we use the parametric DEM technique developed by \citet{Nuevo2015}, adapted for the specific study of bright AR loops. While the DEM function can be obtained using Markov chain Monte Carlo (MCMC) methods \citep{mcmc} or regularised inversion techniques \citep{hannahkontar} for spectroscopic data, the parametric modelling of the DEM is more adequate when narrow-band images are used as input, because it gives simple and computationally efficient solutions.
However, recent works have shown improvements of MCMC DEM modelling applied to AIA narrow-band images using the six channels simultaneously \citep{Schmelz2016,Schmelz2015}. 
Here, we use the  131, 171, and 193 \AA\ AIA channels, where the studied loops {exhibit a larger S/N ratio}. 

{The intrinsic intensity amplitude $I^{(0)}_{k,i}$ determined at each analysed location, of each channel can be written in terms of the ${\rm DEM}(T)$ function at that location as}
\begin{equation}
 I^{(0)}_{k,i} = \int dT \, {\rm TRF}_k(T)\, {\rm DEM}_i (T) \,,
\end{equation}
where ${\rm TRF}_k(T)$ is the temperature response function, TRF, of channel $k$. We compute the TRFs of AIA channels using the plasma emission model CHIANTI 7.1 \citep{Landi2013}. For the calculation of the model we use the abundance set {\tt sun\_coronal\_2012\_schmelz.abund} \citep{Schmelz2012} and the ionization equilibrium calculation set {\tt chianti.ioneq} \citep{Bryans2009}. 

We model the DEM as a Gaussian function with three free parameters: amplitude, centroid, and width, and we obtain the solution finding the values of the parameters that best reproduce the intensities $I^{(0)}_{k,i}$. 

Once the DEM is determined {at each location $s_i$}, the emission measure, EM, and the mean electron temperature across the loop section, $T_m$, can be computed by taking its zeroth and first moments,
\begin{equation}
{\rm EM}_i =\int dT \, {\rm DEM}_i(T) \,, 
\end{equation}
\begin{equation}
 T_{m,i} =\int dT \,T\, {\rm DEM}_i(T) / {\rm EM}_i \,.
\end{equation}
Assuming a circular cross-section, the diameter of the loop {at each location $s_i$} can be estimated as the full width at half maximum (FWHM) of the Gaussian function in Equation \ref{gauss+lin}, $d_i=2\sqrt{2 {\rm ln}2}\,\sigma_{0,i} \approx 2.35\, \sigma_{0,i}$ \citep{Aschwanden2013}. {Other methods can be used} to estimate the diameter of the loop. For example, \citet{LopezFuentes2006} used the relation between the standard deviation of the cross-section intensity profile and the radius of the loop (after background subtraction). In {that} work, the authors showed that both methods lead to similar loop-width values. 

The electron density {at each location $s_i$} can be obtained from the ${\rm EM}_i$ and the diameter $d_i$ of the loop using the following expression \citep{Aschwanden2013}:
\begin{equation}
 N_{e,i} = \sqrt{\frac{{\rm EM}_i}{f\,d_i}} \,,
\end{equation}
where $f$ is the filling factor, i.e. the fraction of the observed volume occupied by the emitting plasma. In this work we assume $f=1$, so the values of the reconstructed electron density are a lower limit. As an example, if $f=0.75$,  the values of the reconstructed electron density increase {by} around 15\%.  

The second moment of the DEM is the standard deviation of the thermal distribution described by the DEM.
\begin{equation}
 W_{T,i}^2 = \int dT (T - T_{m,i})^2 \, {\rm DEM}_i(T)/ {\rm EM}_i \,.
\end{equation}
$W_{T,i}$ is a measure of the {multithermality of the plasma across the loop section at each location $s_i$}.
  
%%%%%%%%%%%%%%%%%FIELD MODELING%%%%%%%%%%%%%%%%%%%%%%%%%%%%%%%%%%%%%%%%

\subsection{Magnetic Field Modeling}  
\label{Bfield_sec}

To obtain the three-dimensional (3D) geometry of the observed loop, we model the magnetic field using the LFFF approximation, given by
\begin{equation}
 \curl \bb = \alpha \,\bb \,,
\end{equation}
where $\alpha$ is a constant that is the free parameter of the model. 
We compute the model using the discrete fast-Fourier transform method discussed by \citet{Alissandrakis1981} and its implementation, including the transformation of coordinates from the local AR frames to the observed ones, by \citet{Mandrini96} and \citet{Demoulin97}.
Since the magnetic configurations of all the extrapolated ARs have a low shear,
a LFFF model is good enough to represent the analysed loops.

We use as boundary condition the line-of-sight magnetogram obtained with HMI closest in time to the 171 \AA\ AIA image from which each target loop is identified. 
In the top panels of Figure \ref{EUVfig}, we show in red the projection on the image of the magnetic field line of the corresponding LFFF model that better matches the shape of the selected EUV bright loop. 
In the bottom panels of the same figure we depict the modelled field lines using another point of view to illustrate their 3D geometry. 

We set the value of $\alpha$ that best matches the observed loop following the procedure discussed by \citet{Green2002}.  
This procedure lets us assign the approximate 3D coordinates to the sample locations where the DEM analysis is performed (black dots in Figure \ref{EUVfig}).

%%%%%%%%%%%%%%%%%%%%%%RESULTS%%%%%%%%%%%%%%%%%%%%%%%%%%%%%%%%%%%%%%%%%%%%%%

\section{Results}
\label{res}

\subsection{Reconstructed Plasma Parameters of the Loops}
\label{recon_par}

\begin{figure}
  \centering
  \includegraphics[width=\textwidth]{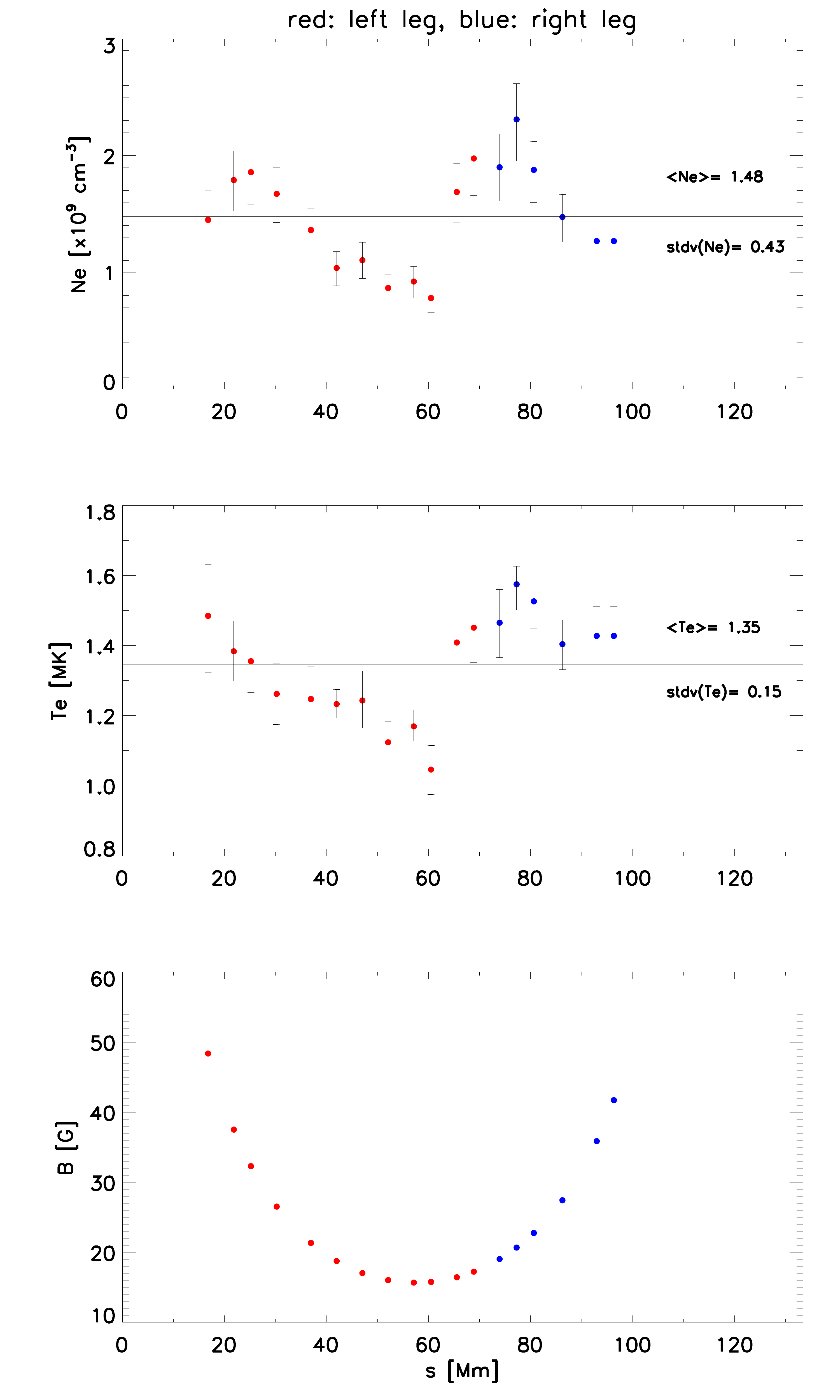}
  \caption{{\bf Variation of $N_e$, $T_e$ and $B$ as a function of the distance $s$ along the loop measured from the photosphere, for loop number 1. Each leg (the loop segment between the photosphere and its apex) is plotted in different colour.  In the top (middle) panels the value of the mean and standard deviation of the density (temperature) along the loop is shown. The error bars in the density and temperature data points are computed as described in Section \ref{recon_par}.}}
  \label{profiles_loop1}
\end{figure}

\begin{figure}
  \centering
    \includegraphics[width=\textwidth]{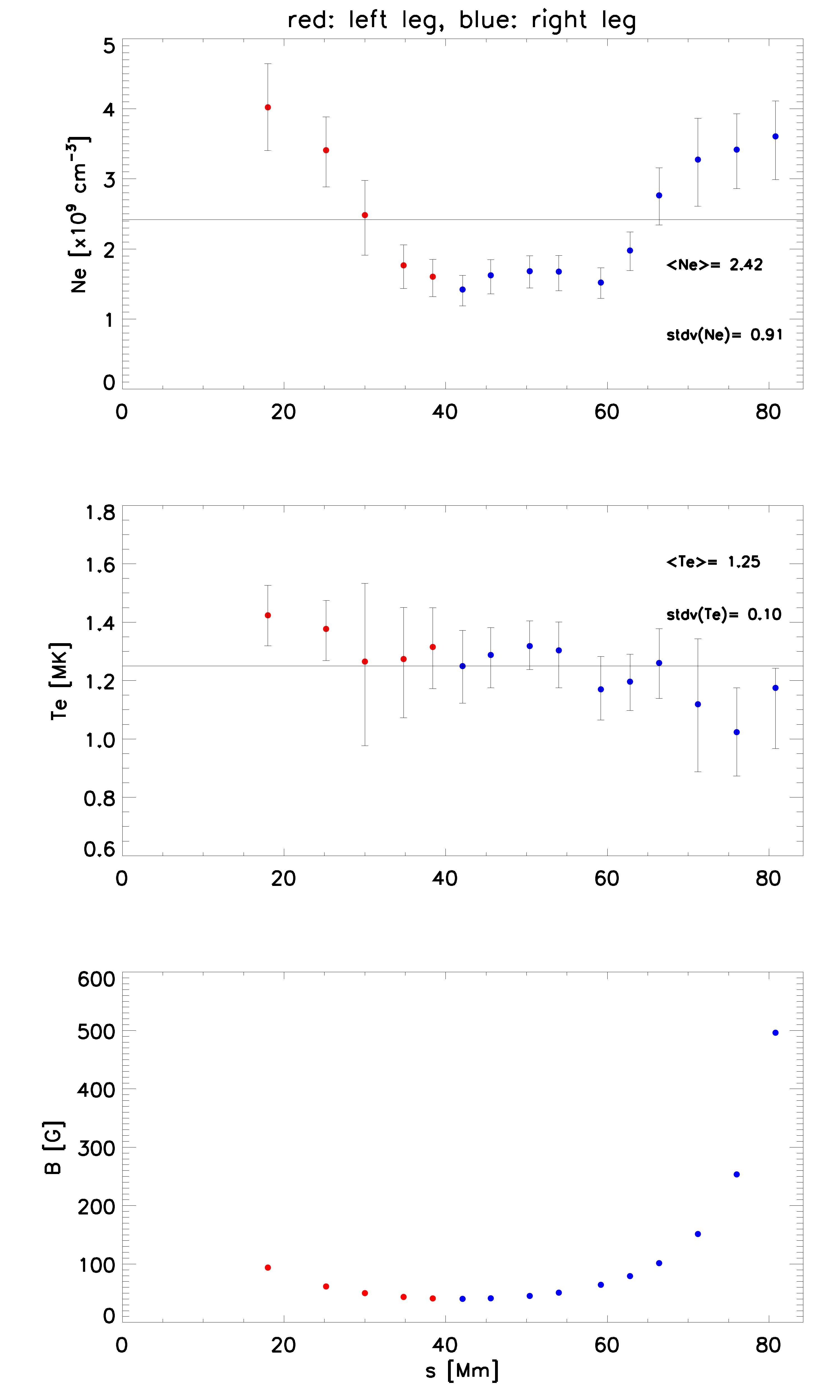}
  \caption{Same as Figure \ref{profiles_loop1} for loop number 6.}
  \label{profiles_loop6}
\end{figure}

Combination of the density and temperature values, determined from the DEM analysis, with the geometry of the loops, reconstructed using LFFF extrapolations, provide a description of the 3D distributions of the physical parameters of the loops. For example, Figures \ref{profiles_loop1} and \ref{profiles_loop6}  show the variation with the distance $s$ along the loop (measured from the photosphere) of the reconstructed electron density, $N_e$, mean temperature, $T_m$, and magnetic field strength, $B$, for loops number 1 and 6, respectively.

Figures \ref{profiles_loop1} and \ref{profiles_loop6} display error bars for the electron density and mean temperature, which depend on the uncertainty of the intensities used for the DEM analysis. The dominating sources of uncertainty are the radiometric calibration of the AIA filters, and the fit and background subtraction procedure described in Section \ref{DEM_sec}. The absolute radiometric calibration uncertainty is of the order of 25\% and the relative calibration uncertainty among the different channels is of the order of 10\% \citep{boerner14}. The uncertainty introduced by the fit and background subtraction procedure is estimated in this work using Equation \ref{back_error}. 

To estimate the error bars of $N_e$ and $T_m$, $\Delta N_e$ and $\Delta T_m$, we perform an ``error box'' analysis, considering simultaneously both sources of uncertainty in a similar fashion to the method described in detail in \citet{Nuevo2015}. Treating uncertainties as independent, we vary the observed trios of EUV intensities used for the DEM analysis, either increasing or decreasing each of the three observed values, considering all possible combinations. In this way, $2^3=8$ DEM analysis are performed at each sample point of every loop. As a result, at each point of every loop eight possible values of the electron density and temperature are obtained. The error bar in density (or temperature) is estimated using the corresponding maximum and minimum values.

\begin{table}
\begin{center}
 \begin{tabular}{cccccccc}
\hline
 Loop  &  $\langle N_e \rangle$ & ${\rm Stdv}(N_e)$ &  $\langle T_m \rangle$  & ${\rm Stdv}(T_m)$ & $\langle \frac{W_T}{T_m} \rangle$ & $\langle \frac{\Delta T_m}{T_m} \rangle$ & $L$ \\
 number   & [$10^9$ cm$^{-3}$]  &  [$10^9$ cm$^{-3}$] & [MK] & [MK] &  &  & [Mm] \\      
 \hline
 1 & 1.48  & 0.43  & 1.35  & 0.15 & 0.25 & 0.08 & 137\\
 2 & 1.68  & 0.62  & 1.25  & 0.14 & 0.22 & 0.07 &  176\\
 3 & 1.86  & 0.45  & 1.12  & 0.15 & 0.36 & 0.08 & 128 \\
 4 & 1.29  & 0.22  & 1.17  & 0.21 & 0.38 & 0.08 & 121 \\
 5 & 1.65  & 0.47  & 1.56  & 0.23 & 0.15 & 0.05 & 208 \\
 6 & 2.42  & 0.91  & 1.25  & 0.10 & 0.37 & 0.11 & 84   \\       
 7 & 1.19  & 0.22  & 1.26  & 0.09 & 0.15 & 0.08 & 93   \\
 8 & 1.40  & 0.41  & 1.34  & 0.09 & 0.26 & 0.12 & 132 \\   
 9 & 1.31  & 0.53  & 1.13  & 0.19 & 0.34 & 0.10 & 337 \\
 10 & 0.56 & 0.15  & 1.08  & 0.16 & 0.23 & 0.10 & 195 \\
 11 & 0.84 & 0.43  & 1.05  & 0.20 & 0.15 & 0.08 & 353  \\
 12 & 1.03 & 0.31  & 1.14  & 0.21 & 0.33 & 0.10 & 175 \\
 13 & 1.28 & 0.60  & 1.10  & 0.23 & 0.31 & 0.10 & 163 \\
\hline
\end{tabular}
\end{center}
\caption{For each analysed loop (rows), columns 2 to 5 report the loop-averaged, $\langle  \rangle$, and standard deviation, Stdv, value of the electron density and mean temperature derived from the computed DEM. Column 6 reports the loop-averaged value of the ratio of the thermal width to the mean temperature. Column 7 shows the loop-averaged value  of the ratio of the  uncertainty of the mean temperature to the mean temperature calculated using the computed DEM. Column 8 reports the length of the loop from the LFFF model.}
\label{table_NyT}
\end{table}

For each analysed loop, we characterise the results of the DEM analysis computing the average value $\langle q \rangle$ and standard deviation $\rm{StDev}(q)$ of each DEM product $q= N_e, T_m,$ and $W_T$. 
%For each analysed loop, we characterise the results of the DEM analysis computing the loop-average value, $\langle  \rangle$, and standard deviation, Stdv, of each DEM product, $N_e$, $T_m$ and $W_T$. 
Table \ref{table_NyT} summarizes the results for all 13 target loops. For all loops, the $\langle N_e \rangle$ values are within the range $0.5\,-\,2.5$ $\times 10^9$ cm$^{-3}$, while the $\langle T_m \rangle$ values are in the range $1.0-1.5~\rm{MK}$. These values are consistent with the characteristic ones obtained in previous studies of bright EUV loops observed in ARs \citep{Aschwanden2011,Aschwanden2013}. 
{Also, for all loops studied in this work, the cross-sectional DEM average temperature $T_m$ does not vary significantly along their main axis, with the DEM temperature standard deviation in the range $0.05-0.2 \langle T_m \rangle$.}
%Also, for the loops studied in this work, the temperature (the cross-sectional averaged value) does not vary significantly along their main axis, with the temperature standard deviation ranging from 5 to 20\% of $\langle T_m \rangle$.

Column 6 of Table \ref{table_NyT} shows the loop-averaged ratio between the thermal width, $W_T$, and the mean temperature $T_m$. Column 7 shows the loop-averaged ratio between the temperature uncertainty $\Delta T_m$ and $T_m$.
Comparing $\langle W_T/T_m \rangle$ with  $\langle \Delta T_m/T_m \rangle$, we consider that the studied loops are not isothermal, globally speaking, but multithermal structures. However, as this thermal width holds all along our loops, we can assure that their temperature, which is an average on the cross-sectional thermal distribution, does not substantially vary along the loop. 

%%%%%%%%%%%%%%%%%%%%%%%TEMPORAL EVOLUTION%%%%%%%%%%%%%%%%%%%%%%%%%%%%%%%%%%%%%%%%

\subsection{Temporal Evolution of Loop Parameters: A Case Study}
\label{S-evol_temp}

To study the temporal evolution of the computed plasma parameters we take loop number 1 as case study. We analyse the loop over a period of two hours and an a half (19:16 to 21:46 UT) from the first to the last AIA 171 \AA\ images in which it is clearly identifiable. For each observation time we reconstruct the density and temperature profiles using the technique described in Section \ref{DEM_sec}. Before 19:16 UT and after 21:46 UT the loop is very diffuse, so that the fractional uncertainty introduced by the fit and background subtraction procedure is of order one. 
\begin{figure}
\centering
{\includegraphics[width=0.78\linewidth]{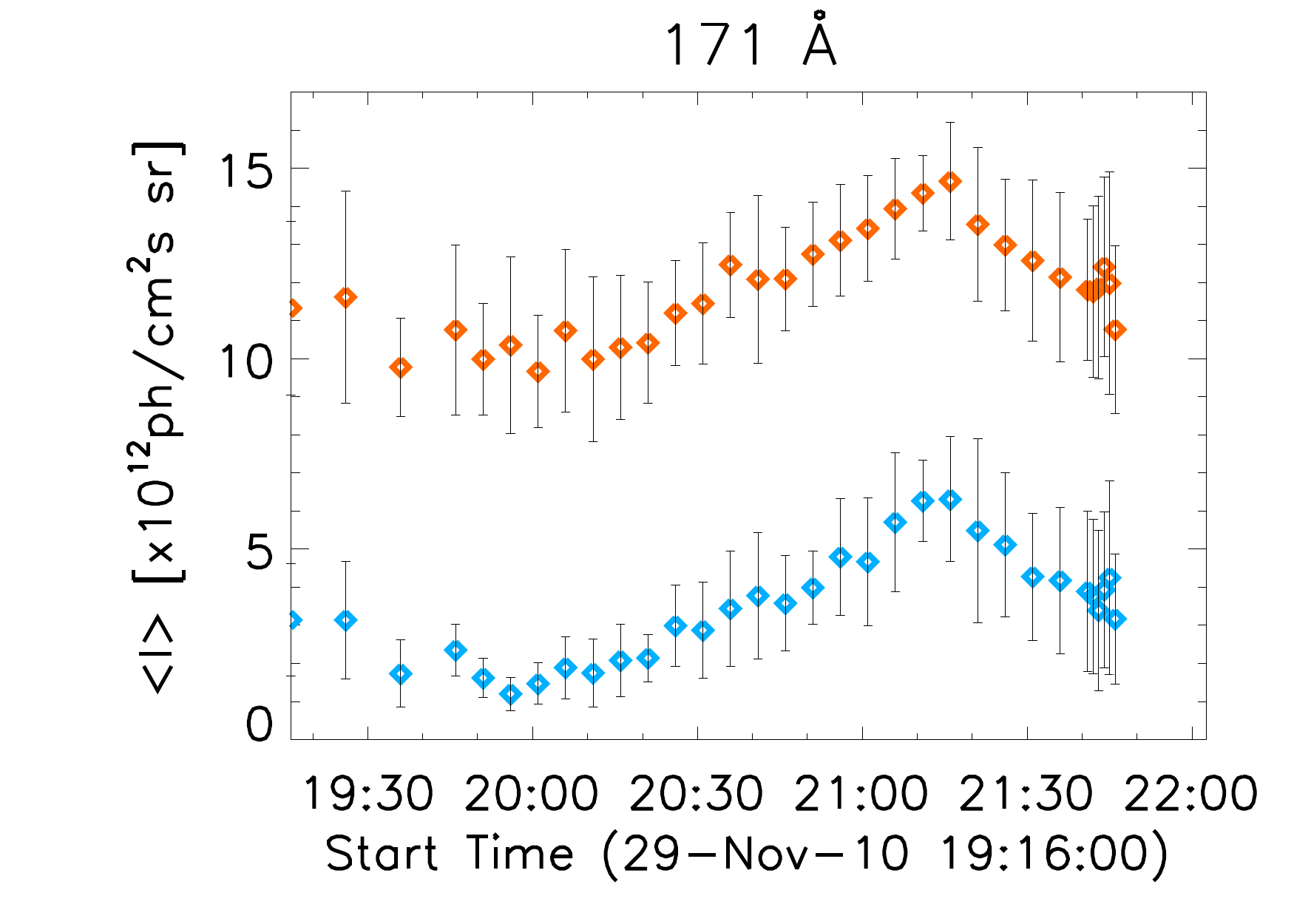}}
{\includegraphics[width=0.78\linewidth]{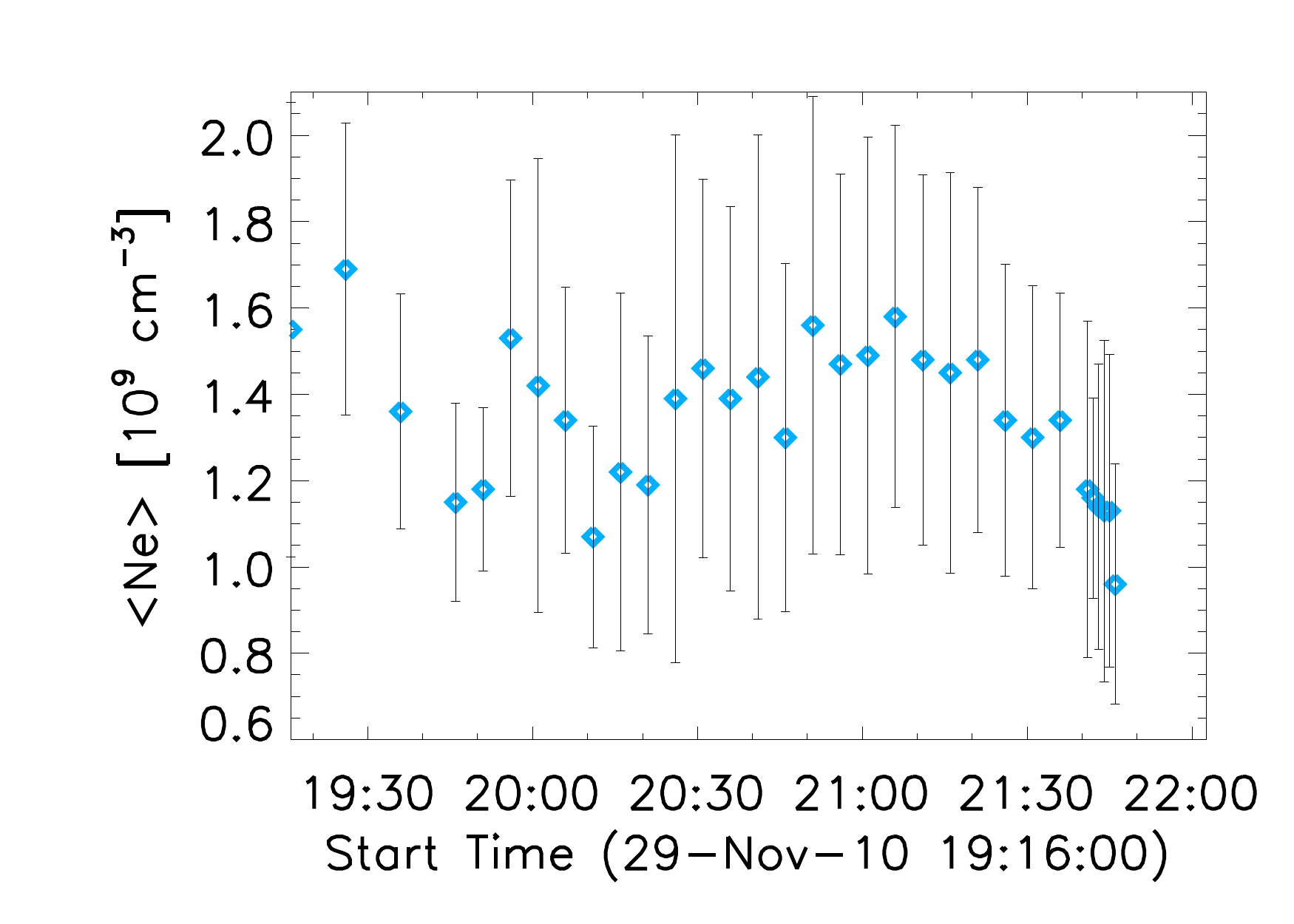}}
{\includegraphics[width=0.78\linewidth]{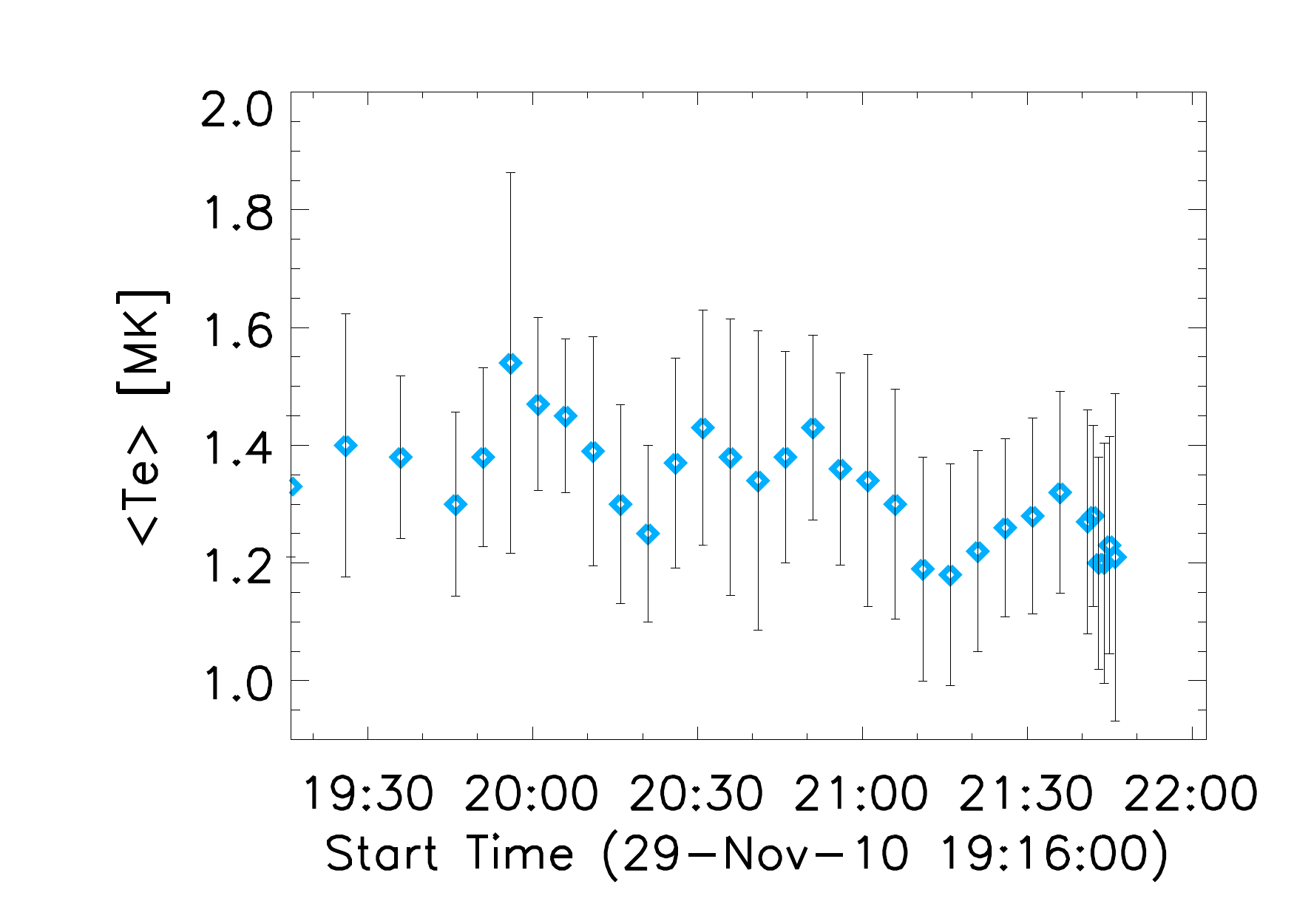}}
\caption{For loop number 1, the diamonds in each panel indicate the temporal evolution of a loop-averaged quantity, namely, the 171 \AA\ intensity (top panel), the DEM electron density (middle panel), and the DEM mean temperature (bottom panel). In the top panel, the intensity is shown before and after background subtraction in orange and blue colours, respectively. In the middle and bottom panels blue colour is used as derived from the DEM analysis based on the background subtracted intensities. At each time, the error bars indicate the standard deviation of all measured values along the loop.}
\label{evol_figs}
 \end{figure}
Figure \ref{evol_figs} shows the loop-averaged values of the 171 \AA\ intensity, electron density, and temperature at each time. For each data point, the bar indicates the standard deviation of the values along the loop at that time. In the top panel the orange dots are the 171 \AA\ intensities without background subtraction and the blue dots are the values with background subtraction. We find that the background emission at each location along the loop accounts for $70\,-\,90$ \% of the measured total intensity.

For this loop, density and temperature do not vary in time in a significant way within the standard deviation values. The results for the rest of the loops in our set are similar. This suggests that all the analysed loops detected in the AIA 171 \AA\ channel are at their mid-lifetime period ({i.e.} not rising nor decaying in intensity) and therefore the obtained parameters are typical of these loops.

%%%%%%%%%%%%%%%%%%COMPARISON WITH MODELS%%%%%%%%%%%%%%%%%%%%%%%%%%%%%%%%%%%%%%%%

\section{Comparison of the Results with HD Models}
\label{ebtel_sec}

\subsection{Constant Heating}
\label{const_heat}

To characterise the thermodynamic state of the observed loops, we use the results of the DEM analysis to constrain hydrodynamic (HD) numerical simulations of the loops.
We compute equilibrium solutions of the zero-dimensional (0D) HD model {Enthalpy-Based Thermal Evolution of Loops} (EBTEL), developed by \citet{Klimchuk2008}. The EBTEL model provides the temporal evolution of the plasma parameters averaged along the loop. 
This model divides the loop in two segments, one corresponding to the corona and the other to the transition region (TR). It also assumes that the loop is symmetric with a uniform cross-section. Integrating the energy balance equation between the coronal base of the loop and its apex, the following expression is obtained
\begin{equation}
 \frac{3L}{4}\dpar{{\langle P \rangle}_{\rm E}}{t} \approx \frac{5}{2}P_0\,V_0 +F_0 + \frac{L}{2}E_H -\phi_{R,{\rm E}} \,,
 \label{balance}
\end{equation}
where $\phi_{R,{\rm E}}=\langle N_e \rangle^2_{\rm E}\,\Lambda \left(\langle T_e \rangle_{\rm E}\right)$ is the radiative loss flux in units of [erg cm$^{-2}$ s$^{-1}$] and $\Lambda \left(\langle T_e \rangle_{\rm E}\right)$ is the radiative loss function. The first two terms on the right-hand side of Equation \ref{balance} are the enthalpy and conductive flux at the coronal base, respectively. The quantities ${\langle P \rangle}_{\rm E}$, ${\langle N_e \rangle}_{\rm E}$, and ${\langle T_e \rangle}_{\rm E}$ are the loop-averaged values of the pressure, density, and temperature, respectively. The quantity $E_H$ is the heating rate in units of [erg cm$^{-3}$ s$^{-1}$], assumed here to be uniform along the loop and constant in time, and $L$ is the length of the loop.
Integrating the energy balance equation in the transition region, a similar equation can be derived.

The energy balance equations of the corona and the TR can be combined to solve the temporal evolution of ${\langle P \rangle}_{\rm E}$, ${\langle N_e \rangle}_{\rm E}$, and ${\langle T_e \rangle}_{\rm E}$.

In our simulations the free parameters of the EBTEL model are the length of the loop, $L$, and the constant heating rate, $E_H$. 
For each of the thirteen studied loops we found the value of $E_H$ for an equilibrium solution of the EBTEL model with the loop length $L$ obtained from the LFFF model (column 6 of Table \ref{table_NyT}) and the loop-averaged temperature value from the DEM analysis, specifically from column 4 of Table \ref{table_NyT}. The last column of Table \ref{VH} shows the ratio between the loop-averaged density value, $\langle N_e \rangle$, of the DEM analysis (column 3 in Table \ref{table_NyT}) and the loop-averaged density predicted by EBTEL assuming a constant heating rate, $\langle N_e \rangle_{\rm E_{\rm const}}$.

The average density values obtained with the DEM analysis are 5 to 150 times larger than those predicted by the equilibrium solution of the EBTEL model. These results strongly suggest that the studied loops are overdense as is the case for warm loops studied in other works \citep{Aschwanden2001,Winebarger2003}. 
%Furthermore, the overdensity factors are consistent with previous studies of EUV loops using TRACE data \citep{Aschwanden2001,Winebarger2003}.

One of the main differences between warm and hot loops is the dominant cooling mechanism at their respective characteristic temperatures. While warm loops are primarily cooled by radiative losses, hot loops are mainly cooled by thermal conduction.
\citet{LopezFuentes2007} derived the following scaling laws for the characteristic times associated to the processes of radiative and thermal conduction cooling:
\begin{equation}
 \tau_{\rm cond} \approx \frac{21}{8}\frac{k_B}{\kappa_0}\frac{N_e\,L^2}{T_e^{5/2}} \,,
 \label{t_cond}
\end{equation}
\begin{equation}
 \tau_{\rm rad} \approx \frac{3 \, k_B}{\Lambda(T_e)}\frac{T_e}{N_e} \,,
 \label{t_rad}
\end{equation}
where $\kappa_0$ is the Spitzer thermal conductivity coefficient and $k_B$ is the Boltzmann constant.
To estimate the characteristic cooling time of each loop, the values of $N_e$ and $T_e$, determined by the DEM analysis, and the loop length $L$ are used.

\begin{figure}	
  \centering
  \includegraphics[width=0.8\textwidth]{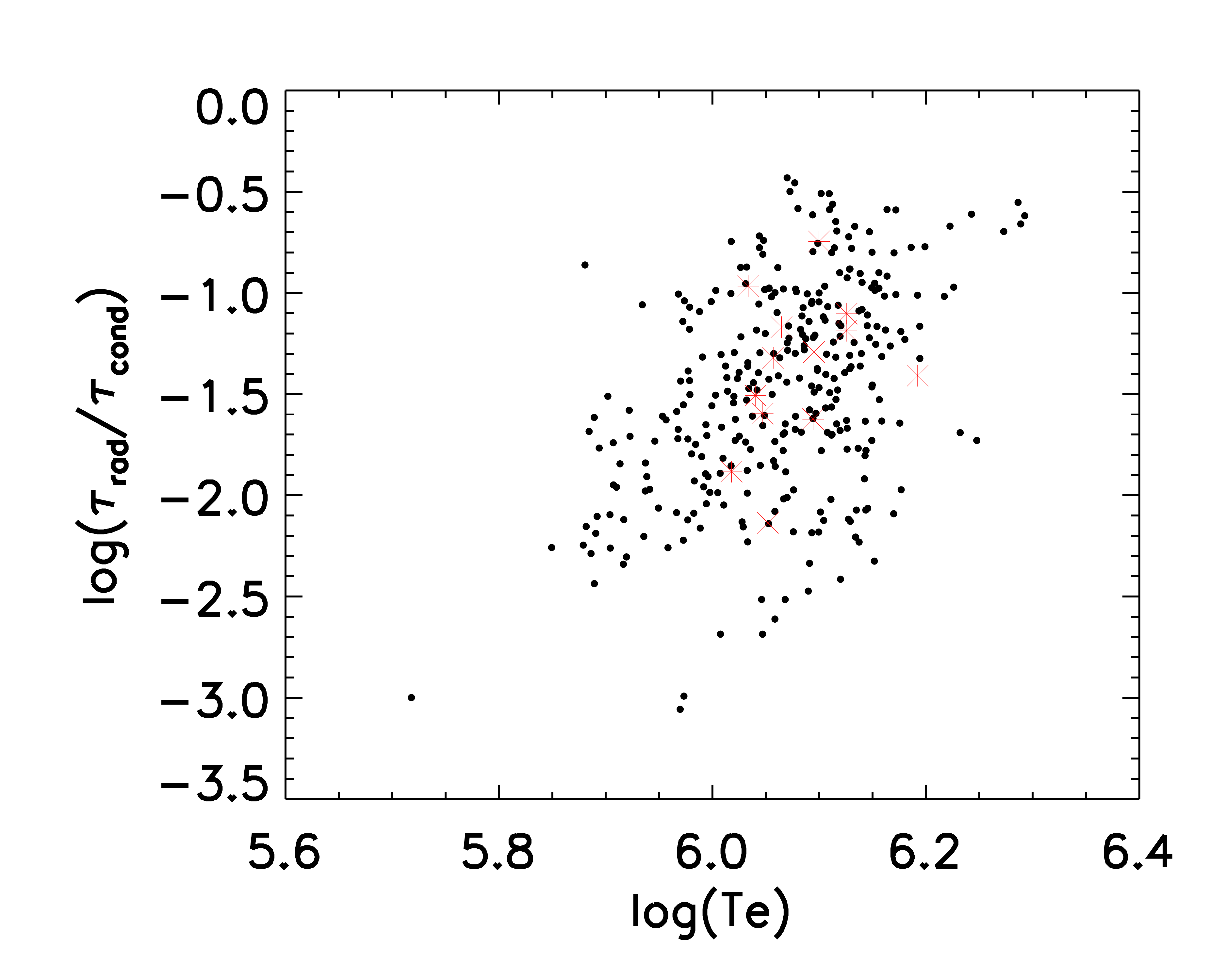}
  \caption{Scatter plot of the ratio $\tau_{\rm rad}/\tau_{\rm cond}$ versus $T_e$ in logarithmic scale. Each black dot corresponds to each point analysed along every loop. Red asterisks show the loop-averaged result for each loop.}
  \label{cooling_times}
\end{figure}

Figure  \ref{cooling_times} shows a scatter plot of the ratio of cooling times versus temperature for all the data points along the thirteen studied loops.
The values of the ratio $\tau_{\rm rad}/\tau_{\rm cond}$ in Figure \ref{cooling_times} can be directly compared with those shown in Figure 6 of \citet{LopezFuentes2007}. This comparison shows that the characteristic cooling times obtained in this work are similar to those found in previous loop observations done with TRACE \citep{Winebarger2003}. 
Indeed, the mean value and standard deviation of $\tau_{\rm rad}/\tau_{\rm cond}$ for the AIA loops studied here are $\approx -1.5$ and $\approx 0.5$, respectively, while for the loops studied with TRACE \citep{Winebarger2003} they are $\approx -1.2$ and $\approx 0.7$, respectively.

Using Equations \ref{t_cond} and \ref{t_rad} we can also compute the characteristic cooling times of the analysed loops, defined as $\tau_{\rm cool} = [ 1/\tau_{\rm cond} + 1/\tau_{\rm rad}]^{-1}$. The computation of this quantity for the loops in our set gives typical values of around 40 min. Notice that this is much shorter than the observed lifetime ($\approx 2.5$ hours) of the loop studied in Section \ref{S-evol_temp}.

%%%%%%%%%%
{
\subsection{Impulsive Heating by Nanoflares}
\label{var_heat}

The results of the previous section indicate that the analysed loops are overdense with respect to equilibrium solutions. As discussed in Section \ref{S-Introduction} these high densities can be expected if the loop plasma evolves in a highly dynamic scenario, as it would be the case if it is heated by short impulsive events or nanoflares \citep[see e.g.][]{cargill_1994}. In the classical nanoflare model \citep{parker_1988} coronal loops are assumed to be formed by subresolution magnetic strands that are individually heated by reconnection events. Observed loops are then the resolution-level manifestation of sets of individual strands evolving more or less independently. The apparently uniform evolution of the loop analysed in Section~\ref{S-evol_temp} is consistent with this scenario. The sum of the radiation produced by several strands evolving impulsively may contribute to the collective appearance of a quasi-static loop.  

To explore this possibility, we use EBTEL to model the evolution of individual strands heated by nanoflares. {If we consider that an observed loop is the collective manifestation of a set of strands, we can assume that at each instant of time the mean thermal properties of the loop approximately coincide with the temporal average of a typical individual strand. Thus, using EBTEL to model the evolution of an individual strand heated by nanoflares we can compare its average coronal temperature and density with the thermal properties of the observed loops, as derived from the DEM analysis. The aim is to show the plausibility of the nanoflare model to explain the high densities observed.}  

As in Section~\ref{const_heat}, the EBTEL inputs are the loop length and the heating rate, not set constant in this case.
The nanoflares are modelled with triangular functions of duration $\tau$, with values in the range $50\,-\,250$~s, and a total volumetric energy in the range $5\,-\,20$~erg cm$^{-3}$. 
Since in the nanoflare model the energy input proceeds from the photospheric motions that inject magnetic energy via reconnection between neighbouring strands, $\tau$ is bounded by the solar granulation characteristic time ($\approx 300~{\rm s}$). Considering a typical subresolution magnetic strand, with a circular cross-section of 100~km diameter and 100~Mm length, the total injected energy lies in the nanoflare range $\approx 10^{-25}\,-\,10^{-24}$~erg. 

Since our objective is to model a standard nanoflare in a typical strand, we trigger the test nanoflare starting from initial conditions that can be considered as characteristic during the evolution. Running a sequence of identical nanoflares separated by a given time interval, we find that after the third nanoflare all following events exhibit identical evolution.
We then use a sequence of three consecutive nanoflares separated by a time interval $\tau_{\rm nf}$ and we use the third one as the test nanoflare. The first two nanoflares are used to create {standard initial conditions} for the strand. 
According to the criterion used to classify nanoflare frequencies as high, intermediate or low, by \citet{lopezfuentes_2015}, if the lapse between consecutive nanoflares is such that the temperature is between 14\% and 67\% of its maximum value, the nanoflare frequency is considered intermediate. The nanoflare frequency in our simulations is defined by the time interval $\tau_{\rm nf}$. We set it by forcing that the temperature of the strand does not decrease more than an 86\% of its maximum value. Thus, the loop is not significantly cooled before the occurrence of the next nanoflare and we obtain frequency values in the intermediate range.

To perform the DEM parametric technique we have used a set of EUV filters with specific temperature sensitivity ranges. Inspecting the response functions of the AIA channels \citep[see ][their Figure 13]{lemen_2012}, we consider the maximum response to be in the temperature range 0.03\,--\,2.27 MK.  To be observable by the three AIA channels, the range of temperatures of the coronal plasma must significantly overlap this range. Therefore, we let the third nanoflare (the last one) to cool down over a time lapse $t_{\rm evol}$, until the loop reaches temperatures within that range. Then, we compute the mean temperature and density of the plasma during the time it remains in that range. We consider that the temperature values determined from the DEM analysis represent the average plasma conditions normally present in the analysed loops when they are detected in the AIA channels.

\begin{figure}[ht]
\begin{center}
\includegraphics[width=0.90\textwidth]{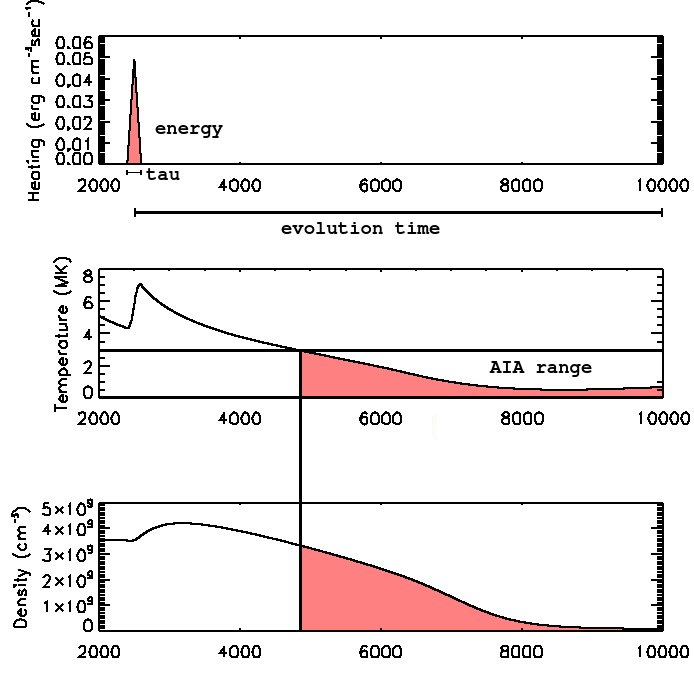}
\caption{Top panel: Example of the time evolution of the plasma parameters of a magnetic strand with a length of 121 Mm heated by a triangular nanoflare (the third of a series of three) with $\tau=$200 s and a maximum heating rate of 0.05 erg cm$^{-3}$ s$^{-1}$. The middle and bottom panels show, respectively, the evolution of the strand averaged coronal-temperature and density due to the nanoflare. The pink shaded areas indicate the plasma temperature and density within the response range of the used AIA channels.}
\label{ex}
\end{center}
\end{figure}

The top panel of Figure~\ref{ex} shows the temporal evolution of the heating power  of the test nanoflare indicating its duration, $\tau$, and the evolution time, $t_{evol}$. The middle and bottom panels show the temporal evolution of the coronal mean temperature and density over time. We compute the time-average temperature and density along the time the plasma is in the temperature range 0.03\,--\,2.27 MK, indicated {by a pink} shaded area in the middle and bottom panels of the figure.

We apply the procedure discussed above to the loops analysed in the previous sections. We vary the model parameters in order to reproduce the thermal properties of the different observed loops. Table \ref{VH} shows the EBTEL parameters used for the different cases: loop length, $L$; nanoflare energy, $E_{\rm nf}$; nanoflare duration, $\tau$; waiting time between heating events to prepare the initial conditions for the test nanoflare, $\tau_{\rm nf}$; and the evolution time, $t_{\rm evol}$, for the time averaging of the plasma parameters. The last three columns of the table show a comparison of the DEM results with the average density and temperature obtained with EBTEL, using constant and impulsive heating. The results indicate that the model based on nanoflare heating is more consistent with the observations. It can be seen that longer loops need longer integration times and more energy input than shorter ones in order to reproduce their plasma properties.  

\begin{table}
\begin{center}
\begin{tabular}{c c c c c c c c}
  \hline
  %\\
L & $E_{\rm nf}$ & $\tau$ & $\tau_{\rm nf}$ & $t_{\rm evol}$ & $\frac{\langle T_e \rangle}{\langle T_e \rangle_E}$ & $\frac{\langle N_e \rangle}{\langle N_e \rangle_E}$ & $\frac{\langle N_e \rangle}{ \langle N_e \rangle_{E_{\rm const}}}$\\
$[$Mm$]$ & [erg cm$^{-3}$] & [s] & [s] & [s] &  &  & \\
 %\\
   \hline
   \\
 137 &  5 & 200 & 1200 &  6100 & 1.06 & 1.04 &   9.0 \\
 176 &  5 & 250 & 1000 &  6875 & 0.80 & 1.15 &  18.2 \\
 128 &  5 & 200 & 1000 &  5500 & 0.84 & 1.14 &  16.3 \\
 121 &  5 & 200 & 1000 &  6000 & 1.05 & 0.98 &   9.4 \\
 208 &  5 & 100 &  500 &  7750 & 0.91 & 1.13 &  14.4 \\
  84 &  5 &  50 & 1100 &  3675 & 0.97 & 1.12 &   9.2 \\
  93 &  5 & 200 & 1100 &  5000 & 1.24 & 0.83 &   5.0 \\
 132 &  5 & 200 & 1200 &  5800 & 1.03 & 0.94 &   8.1 \\
 337 & 20 &  50 & 1500 & 11875 & 0.77 & 1.45 &  60.8 \\
 195 &  5 & 200 &  500 & 12500 & 1.00 & 1.02 &  10.3 \\
 353 & 20 &  50 & 1500 & 13875 & 0.77 & 0.36 & 161.2 \\
 175 &  5 & 200 & 1000 &  8500 & 0.96 & 1.06 &  23.5 \\
 163 &  5 & 200 &  800 &  7900 & 0.94 & 1.21 &  28.2 \\ 
 \\
  \hline
\end{tabular}
\end{center}
\caption{EBTEL parameters used to model each studied loop: loop length $L$; nanoflare energy ,$E_{\rm nf}$; $\tau$; $\tau_{\rm nf}$; and $t_{evol}$. For comparison in columns {six and seven}, we show the ratio between average temperatures and densities obtained with the parametric DEM method, $\langle T_e \rangle$ and $\langle N_e \rangle$, and the EBTEL model with nanoflare heating, $\langle T_e \rangle_E$ and $\langle N_e \rangle_E$. The last column compares $\langle N_e \rangle$ with the EBTEL densities obtained using constant heating rates, $ \langle N_e \rangle_{E_{\rm const}}$.}
\label{VH}
\end{table}

\begin{figure}[ht]
\begin{center}
\includegraphics[width=0.90\textwidth]{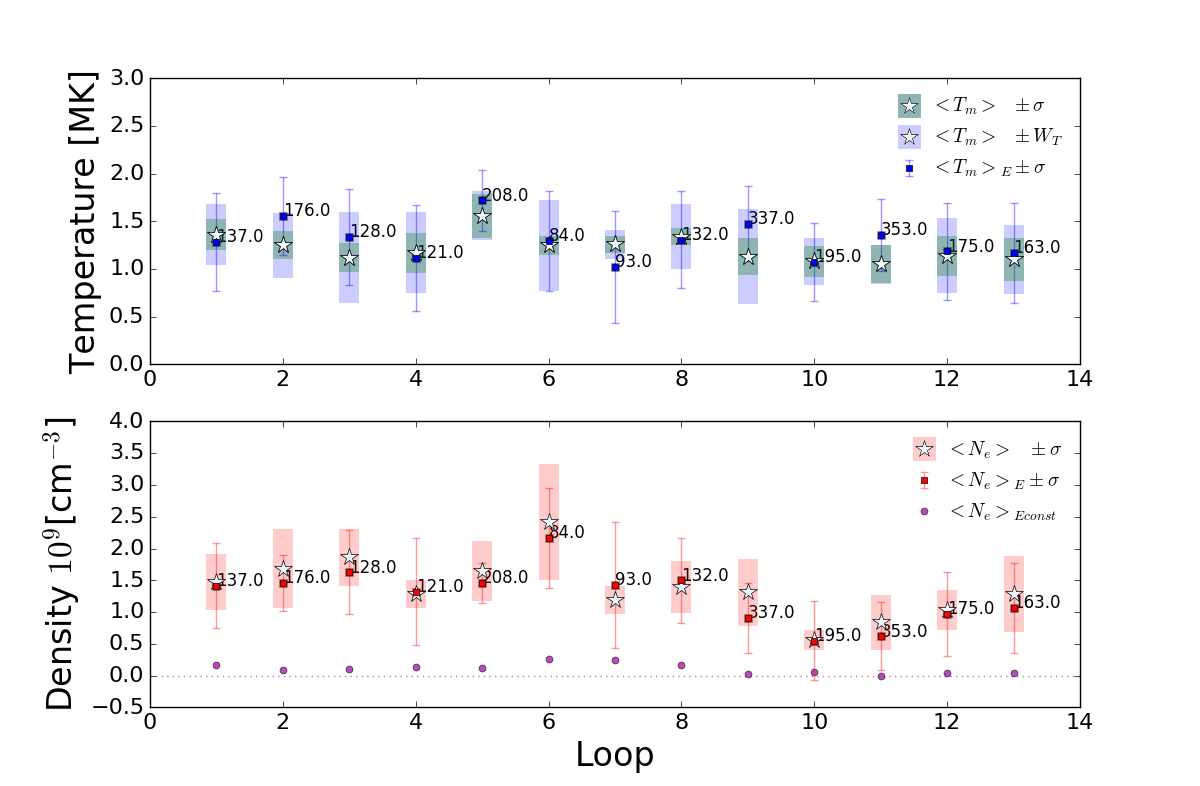}
\caption{Comparison of loop-averaged temperature (upper panel) and density (lower panel) obtained with the parametric DEM method and the mean temperature and density obtained with EBTEL. In the top panels, stars indicate the loop-averaged temperature derived from the DEM. Green and blue areas represent, respectively, the standard deviation and temperature width distribution from the DEM analysis. Blue squares correspond to the time-averaged temperatures obtained with the EBTEL model using impulsive heating. Error bars indicate the standard deviation of the modelled temperatures during the plasma evolution within the AIA response range. The lower panel shows, in pink, the same corresponding means and averages for the density. Violet bullets represent the time-averaged density obtained with EBTEL using a constant heating rate.}
\label{fig}
\end{center}
\end{figure}

For each analysed loop, Figure \ref{fig} shows the loop-averaged temperature (top panel) and density (bottom panel) obtained  with the DEM analysis (stars) and the EBTEL model (squares). In the top panel, green and blue areas represent the standard deviation, $\sigma$, of the DEM electron temperature and the DEM thermal width ,$W_T$, respectively. In the bottom panel, pink areas indicate the standard deviation of the DEM electron density.
The squares correspond to the time-average plasma properties obtained with EBTEL using variable heating. In both panels, the error bars represent the standard deviation of the EBTEL values during the temporal evolution of the simulation. Note that in most cases the standard deviation of the temperature from the EBTEL model is of the same order or larger than $W_T$ of the DEM analysis indicating that the model reproduces the degree of multithermality obtained from the observations with the DEM technique. In the case of density, we also indicate with violet bullets the time-average density obtained with EBTEL using a constant heating rate. 

In most cases we find that the loop-averaged plasma properties obtained from the DEM analysis and with the EBTEL model, using impulsive heating, are consistent. In the case of the density we observe a significant difference between both heating scenarios. Constant heating produces densities one or two orders of magnitude smaller than those obtained with heating by nanoflares. Then, we conclude that the second scenario, heating by impulsive events, better reproduces the plasma properties of the loops analysed here.}

%%%%%%%%%%%%%%%%%%%%%%%%%CONCLUSIONS%%%%%%%%%%%%%%%%%%%%%%%%%%%%%%%%%%%

\section{Conclusions}
\label{S-Summary}

We identify and analyse 13 AR coronal loops observed with SDO/AIA around the time of their maximum brightness. We develop and apply a technique that reconstructs the electron density and temperature distribution along the loops. We combine parametric DEM analysis, including background subtraction, with LFFF extrapolations of the photospheric magnetic field to determine the 3D geometry of the loops.
{Other existing DEM methods that work with multi-channel EUV image sets (see e.g. \citealt{chueng2015}, or \citealt{delzanna2018}, for a review of EUV diagnostic methods) are designed to produce 2D DEM maps.}

{The procedure developed in this work is designed to specifically study EUV bright loops, enabling the background subtraction in a practical and intuitive way. This procedure provides the intrinsic emission of the loop, from which we can derive the local DEM at each point along a bright loop. Combined with 3D magnetic field extrapolations, the method allows reconstructing the 3D distribution of the electron density and temperature along EUV bright loops. It is worth to note that the background emission can contribute with as much as 70~\% of the total observed intensity at a pixel of a EUV bright loop
(see, e.g., \citealt{LopezFuentes2006}, \citealt{Aschwanden2011}). Since we are just interested in the intrinsic loop emission and not in DEM distributions over larger active region areas, we consider that the method used here is more appropriate for this kind of analysis.}

{
In this work we use a LFFF model to extrapolate the coronal magnetic field. A non-linear force free field (NLFFF) model, which provides a better representation of small-scale highly twisted structures, can be also used with our method. However, in all the examples we have studied, loops are of intermediate scale-size and very low twist. Under these conditions, LFFF and NLFFF models provide very similar results (see, e.g, \citealt{Mandrini2014}) with the advantage that a LFFF extrapolation is computationally much less demanding and much faster.  The latter is particularly convenient since our aim is to model a single particular observed loop, and to do so we need to run several models with different $\alpha$ values until we obtain the best match between a computed field line and the observed structure.} 

We obtain error estimations, $\Delta T_m$, based on the effect of the background subtraction on the density and temperature. As an indicator of the temperature inhomogeneity of the loops {(i.e. their multithermality)} we consider the width, $W_T$, of the thermal distribution associated to the DEM. We find that $W_T$ is systematically larger than $\Delta T_m$  and conclude that the studied loops are multithermal{; This suggests} that the loops are formed by strands or sub-loops with different temperatures. We also find that the mean of the temperature distribution on each analysed loop point, $T_m$, does not vary significantly along the loop, being its standard deviation $5\,-\,20$\% of its mean, $\langle T_m \rangle$. {Furthermore, $\langle T_m \rangle$ is comparable to $W_T$, whose mean is 0.32 MK for all the analysed loop points.  This implies  that the temperature variation along each loop and among loops is comparable to the thermal amplitude associated to the DEM distribution.}

For one of the loops in our set, observed in AR 11130, we perform a temporal evolution analysis by applying the DEM parametric method on all the AIA images along the time during which the loop is observable, i.e. for all images in which the loop intrinsic intensity is high enough to distinguish it from the background. Once again, we find that the density and  temperature do not vary substantially in time. The reconstructed values of $N_e$ and $T_e$ along the loop evolution are similar to the instantaneous values obtained for the rest of the loops in our set. This suggests {that all loops have been observed at a similar stage of their evolution, which is their mid lifetime,} 
and that the obtained $N_e$ and $T_e$ values are typical of EUV loops observed with SDO/AIA.

It is of course not surprising that these loops are in the temperature range in which the AIA channels have their maximum response; however, what makes them interesting regarding the structure and dynamics of the solar corona is their apparent spatial coherence along times that are longer than the characteristic cooling times. 
The question arises if the loops are constantly heated at these temperatures during their evolution or if they are the manifestation of processes involving broader temperature ranges, i.e. if they represent the cooling phase of structures evolving from hotter temperatures.
%
%Time lags indicating the presence of cooling processes in SDO/AIA loops have been previously analysed by \citet{viall12}.

In order to test the previous possibilities, we compare our results with the hydrodynamic model EBTEL \citep{Klimchuk2008}. First, we consider that the loops are in quasi-static equilibrium and apply a constant heating producing temperatures consistent with our observations. We find that in this case the predicted densities are too low compared with the observations. As we discussed in Section 1, previous studies have shown that EUV loops are overdense with respect to quasi-static equilibrium solutions \citep{Aschwanden2001,Winebarger2003}. {This overdensity} can be explained if the heating mechanism is impulsive. However, impulsive heating at the footpoints predicts loops in thermal non-equilibrium with important density asymmetries and temperature gradients \citep{Klimchuk2010,Lionello2013}. This is not the case of the loops studied here. It has been proposed that multithermal and overdense loops would be more consistent with impulsive heating along the coronal part of the loop (e.g. in the form of nanoflares, see \citealt{klimchuk15}). To test this possibility we model, for each of the observed loops, the evolution of single strands heated by nanoflares. We find densities and temperatures that reproduce the observed values. We conclude that this heating scenario {explains fairly well} all the characteristics of the plasma in the observed loops: density values, temperature distributions, and temporal evolution. 

\begin{acks}
FAN, MLF, CMacC, CHM, and AMV  acknowledge financial support from the Argentine grants PICT 2016-0221 (ANPCyT) and UBACyT 20020170100611BA.  AMV also acknowledges UBACyT grant 20020160100072BA that partially supported his participation in this research. FAN, MLF, CHM, and AMV are members of the Carrera del Investigador Científico of the Consejo Nacional de Investigaciones Científicas y Técnicas (CONICET). CMacC is a fellow of CONICET. The authors acknowledge the use of data from the SDO (NASA) mission. The authors thank the reviewer for his/her useful comment and suggestions that have helped to improve this article. 
\end{acks}
\\

\begin{footnotesize}
Disclosure of Potential Conflicts of Interest: The authors declare that they have no conflicts of interest.
\end{footnotesize}

\bibliographystyle{spr-mp-sola}
\bibliography{ms_fanuevo_v3}

\begin{thebibliography}{34}
% BibTex style file: spr-mp-sola.bst (nameyear), 2015-03-09
\ifx\bisbn     \undefined \def\bisbn  #1{ISBN #1}\fi
\ifx\binits    \undefined \def\binits#1{#1}\fi
\ifx\bauthor   \undefined \def\bauthor#1{#1}\fi
\ifx\batitle   \undefined \def\batitle#1{#1}\fi
\ifx\bjtitle   \undefined \def\bjtitle#1{\textit{#1}}\fi
\ifx\bvolume   \undefined \def\bvolume#1{\textbf{#1}}\fi
\ifx\byear     \undefined \def\byear#1{#1}\fi
\ifx\bissue    \undefined \def\bissue#1{#1}\fi
\ifx\bfpage    \undefined \def\bfpage#1{#1}\fi
\ifx\blpage    \undefined \def\blpage #1{#1}\fi
\ifx\burl      \undefined \def\burl#1{\textsf{#1}}\fi
\ifx\href      \undefined \def\href#1#2{\textsf{#2}}\fi
\ifx\betal     \undefined \def\betal{\textit{et al.}}\fi
\ifx\bctitle   \undefined \def\bctitle#1{#1}\fi
\ifx\beditor   \undefined \def\beditor#1{#1}\fi
\ifx\bbtitle   \undefined \def\bbtitle#1{\textit{#1}}\fi
\ifx\bedition  \undefined \def\bedition#1{#1}\fi
\ifx\bseriesno \undefined \def\bseriesno#1{\textbf{#1}}\fi
\ifx\blocation \undefined \def\blocation#1{#1}\fi
\ifx\bsertitle \undefined \def\bsertitle#1{\textit{#1}}\fi
\ifx\bsnm      \undefined \def\bsnm#1{#1}\fi
\ifx\bsuffix   \undefined \def\bsuffix#1{#1}\fi
\ifx\bparticle \undefined \def\bparticle#1{#1}\fi
\ifx\barticle  \undefined \def\barticle#1{}\fi
\ifx\binstitute  \undefined \def\binstitute#1{#1}\fi
\ifx\bpublisher  \undefined \def\bpublisher#1{#1}\fi
\ifx\doiurl    \undefined
  \def\doiurl#1{\href{http://dx.doi.org/#1}{\textsf{DOI}}}\fi
\ifx\arxivurl  \undefined
  \def\arxivurl#1{\href{http://arxiv.org/abs/#1}{\textsf{arXiv}}}\fi
\ifx\adsurl    \undefined
  \def\adsurl#1{\href{http://adsabs.harvard.edu/abs/#1}{\textsf{ADS}}}\fi
\ifx\botherref \undefined \def\botherref#1{}\fi
\ifx\url       \undefined \def\url#1{\textsf{#1}}\fi
\ifx\bchapter  \undefined \def\bchapter#1{}\fi
\ifx\bbook     \undefined \def\bbook#1{}\fi
\ifx\bcomment  \undefined \def\bcomment#1{#1}\fi
\ifx\oauthor   \undefined \def\oauthor#1{#1}\fi
\ifx\citeauthoryear \undefined\def \citeauthoryear#1{#1}\fi
\ifx\endbibitem\undefined \def\endbibitem{}\fi
\ifx\bconflocation  \undefined \def\bconflocation#1{#1} \fi

\bibitem[\protect\citeauthoryear{{Alissandrakis}}{1981}]{Alissandrakis1981}
\begin{barticle}
\bauthor{\bsnm{{Alissandrakis}}, \binits{C.E.}}:
\byear{1981},
\batitle{{On the computation of constant alpha force-free magnetic field}}.
\bjtitle{\aap}
\bvolume{100},
\bfpage{197}.
\adsurl{1981A\%26A...100..197A}.
\end{barticle}
\endbibitem

\bibitem[\protect\citeauthoryear{{Aschwanden} and
  {Boerner}}{2011}]{Aschwanden2011}
\begin{barticle}
\bauthor{\bsnm{{Aschwanden}}, \binits{M.J.}},
\bauthor{\bsnm{{Boerner}}, \binits{P.}}:
\byear{2011},
\batitle{{Solar Corona Loop Studies with the Atmospheric Imaging Assembly. I.
  Cross-sectional Temperature Structure}}.
\bjtitle{\apj}
\bvolume{732},
\bfpage{81}.
\doiurl{10.1088/0004-637X/732/2/81}.
\adsurl{2011ApJ...732...81A}.
\end{barticle}
\endbibitem

\bibitem[\protect\citeauthoryear{{Aschwanden}, {Schrijver}, and
  {Alexander}}{2001}]{Aschwanden2001}
\begin{barticle}
\bauthor{\bsnm{{Aschwanden}}, \binits{M.J.}},
\bauthor{\bsnm{{Schrijver}}, \binits{C.J.}},
\bauthor{\bsnm{{Alexander}}, \binits{D.}}:
\byear{2001},
\batitle{{Modeling of Coronal EUV Loops Observed with TRACE. I. Hydrostatic
  Solutions with Nonuniform Heating}}.
\bjtitle{\apj}
\bvolume{550},
\bfpage{1036}.
\doiurl{10.1086/319796}.
\adsurl{2001ApJ...550.1036A}.
\end{barticle}
\endbibitem

\bibitem[\protect\citeauthoryear{{Aschwanden}
  \textit{et~al.}}{2013}]{Aschwanden2013}
\begin{barticle}
\bauthor{\bsnm{{Aschwanden}}, \binits{M.J.}},
\bauthor{\bsnm{{Boerner}}, \binits{P.}},
\bauthor{\bsnm{{Schrijver}}, \binits{C.J.}},
\bauthor{\bsnm{{Malanushenko}}, \binits{A.}}:
\byear{2013},
\batitle{{Automated Temperature and Emission Measure Analysis of Coronal Loops
  and Active Regions Observed with the Atmospheric Imaging Assembly on the
  Solar Dynamics Observatory (SDO/AIA)}}.
\bjtitle{\solphys}
\bvolume{283},
\bfpage{5}.
\doiurl{10.1007/s11207-011-9876-5}.
\adsurl{2013SoPh..283....5A}.
\end{barticle}
\endbibitem

\bibitem[\protect\citeauthoryear{{Boerner} \textit{et~al.}}{2014}]{boerner14}
\begin{barticle}
\bauthor{\bsnm{{Boerner}}, \binits{P.F.}},
\bauthor{\bsnm{{Testa}}, \binits{P.}},
\bauthor{\bsnm{{Warren}}, \binits{H.}},
\bauthor{\bsnm{{Weber}}, \binits{M.A.}},
\bauthor{\bsnm{{Schrijver}}, \binits{C.J.}}:
\byear{2014},
\batitle{{Photometric and Thermal Cross-calibration of Solar EUV Instruments}}.
\bjtitle{\solphys}
\bvolume{289},
\bfpage{2377}.
\doiurl{10.1007/s11207-013-0452-z}.
\adsurl{2014SoPh..289.2377B}.
\end{barticle}
\endbibitem

\bibitem[\protect\citeauthoryear{{Bryans}, {Landi}, and
  {Savin}}{2009}]{Bryans2009}
\begin{barticle}
\bauthor{\bsnm{{Bryans}}, \binits{P.}},
\bauthor{\bsnm{{Landi}}, \binits{E.}},
\bauthor{\bsnm{{Savin}}, \binits{D.W.}}:
\byear{2009},
\batitle{{A New Approach to Analyzing Solar Coronal Spectra and Updated
  Collisional Ionization Equilibrium Calculations. II. Updated Ionization Rate
  Coefficients}}.
\bjtitle{\apj}
\bvolume{691},
\bfpage{1540}.
\doiurl{10.1088/0004-637X/691/2/1540}.
\adsurl{2009ApJ...691.1540B}.
\end{barticle}
\endbibitem

\bibitem[\protect\citeauthoryear{{Cargill}}{1994}]{cargill_1994}
\begin{barticle}
\bauthor{\bsnm{{Cargill}}, \binits{P.J.}}:
\byear{1994},
\batitle{{Some Implications of the Nanoflare Concept}}.
\bjtitle{\apj}
\bvolume{422},
\bfpage{381}.
\doiurl{10.1086/173733}.
\adsurl{https://ui.adsabs.harvard.edu/abs/1994ApJ...422..381C}.
\end{barticle}
\endbibitem

\bibitem[\protect\citeauthoryear{{Cheung} \textit{et~al.}}{2015}]{chueng2015}
\begin{barticle}
\bauthor{\bsnm{{Cheung}}, \binits{M.C.M.}},
\bauthor{\bsnm{{Boerner}}, \binits{P.}},
\bauthor{\bsnm{{Schrijver}}, \binits{C.J.}},
\bauthor{\bsnm{{Testa}}, \binits{P.}},
\bauthor{\bsnm{{Chen}}, \binits{F.}},
\bauthor{\bsnm{{Peter}}, \binits{H.}},
\bauthor{\bsnm{{Malanushenko}}, \binits{A.}}:
\byear{2015},
\batitle{{Thermal Diagnostics with the Atmospheric Imaging Assembly on board
  the Solar Dynamics Observatory: A Validated Method for Differential Emission
  Measure Inversions}}.
\bjtitle{\apj}
\bvolume{807}(\bissue{2}),
\bfpage{143}.
\doiurl{10.1088/0004-637X/807/2/143}.
\adsurl{https://ui.adsabs.harvard.edu/abs/2015ApJ...807..143C}.
\end{barticle}
\endbibitem

\bibitem[\protect\citeauthoryear{{Del Zanna} and {Mason}}{2018}]{delzanna2018}
\begin{barticle}
\bauthor{\bsnm{{Del Zanna}}, \binits{G.}},
\bauthor{\bsnm{{Mason}}, \binits{H.E.}}:
\byear{2018},
\batitle{{Solar UV and X-ray spectral diagnostics}}.
\bjtitle{Living Reviews in Solar Physics}
\bvolume{15}(\bissue{1}),
\bfpage{5}.
\doiurl{10.1007/s41116-018-0015-3}.
\adsurl{https://ui.adsabs.harvard.edu/abs/2018LRSP...15....5D}.
\end{barticle}
\endbibitem

\bibitem[\protect\citeauthoryear{{D{\'e}moulin}
  \textit{et~al.}}{1997}]{Demoulin97}
\begin{barticle}
\bauthor{\bsnm{{D{\'e}moulin}}, \binits{P.}},
\bauthor{\bsnm{{Bagal{\' a}}}, \binits{L.G.}},
\bauthor{\bsnm{{Mandrini}}, \binits{C.H.}},
\bauthor{\bsnm{{H{\'e}noux}}, \binits{J.C.}},
\bauthor{\bsnm{{Rovira}}, \binits{M.G.}}:
\byear{1997},
\batitle{{Quasi-separatrix layers in solar flares. II. Observed magnetic
  configurations.}}
\bjtitle{\aap}
\bvolume{325},
\bfpage{305}.
\end{barticle}
\endbibitem

\bibitem[\protect\citeauthoryear{{Green} \textit{et~al.}}{2002}]{Green2002}
\begin{barticle}
\bauthor{\bsnm{{Green}}, \binits{L.M.}},
\bauthor{\bsnm{{L{\'o}pez fuentes}}, \binits{M.C.}},
\bauthor{\bsnm{{Mandrini}}, \binits{C.H.}},
\bauthor{\bsnm{{D{\'e}moulin}}, \binits{P.}},
\bauthor{\bsnm{{Van Driel-Gesztelyi}}, \binits{L.}},
\bauthor{\bsnm{{Culhane}}, \binits{J.L.}}:
\byear{2002},
\batitle{{The Magnetic Helicity Budget of a cme-Prolific Active Region}}.
\bjtitle{\solphys}
\bvolume{208},
\bfpage{43}.
\doiurl{10.1023/A:1019658520033}.
\adsurl{http://cdsads.u-strasbg.fr/cgi-bin/nph-bib_query?bibcode=2002SoPh..208...43G&db_key=AST}.
\end{barticle}
\endbibitem

\bibitem[\protect\citeauthoryear{{Hannah} and {Kontar}}{2012}]{hannahkontar}
\begin{barticle}
\bauthor{\bsnm{{Hannah}}, \binits{I.G.}},
\bauthor{\bsnm{{Kontar}}, \binits{E.P.}}:
\byear{2012},
\batitle{{Differential emission measures from the regularized inversion of
  Hinode and SDO data}}.
\bjtitle{\aap}
\bvolume{539},
\bfpage{A146}.
\doiurl{10.1051/0004-6361/201117576}.
\adsurl{2012A\%26A...539A.146H}.
\end{barticle}
\endbibitem

\bibitem[\protect\citeauthoryear{{Kashyap} and {Drake}}{1998}]{mcmc}
\begin{barticle}
\bauthor{\bsnm{{Kashyap}}, \binits{V.}},
\bauthor{\bsnm{{Drake}}, \binits{J.J.}}:
\byear{1998},
\batitle{{Markov-Chain Monte Carlo Reconstruction of Emission Measure
  Distributions: Application to Solar Extreme-Ultraviolet Spectra}}.
\bjtitle{\apj}
\bvolume{503},
\bfpage{450}.
\doiurl{10.1086/305964}.
\adsurl{1998ApJ...503..450K}.
\end{barticle}
\endbibitem

\bibitem[\protect\citeauthoryear{{Klimchuk}}{2015}]{klimchuk15}
\begin{barticle}
\bauthor{\bsnm{{Klimchuk}}, \binits{J.A.}}:
\byear{2015},
\batitle{{Key aspects of coronal heating}}.
\bjtitle{Philosophical Transactions of the Royal Society of London Series A}
\bvolume{373},
\bfpage{20140256}.
\doiurl{10.1098/rsta.2014.0256}.
\adsurl{2015RSPTA.37340256K}.
\end{barticle}
\endbibitem

\bibitem[\protect\citeauthoryear{{Klimchuk}, {Karpen}, and
  {Antiochos}}{2010}]{Klimchuk2010}
\begin{barticle}
\bauthor{\bsnm{{Klimchuk}}, \binits{J.A.}},
\bauthor{\bsnm{{Karpen}}, \binits{J.T.}},
\bauthor{\bsnm{{Antiochos}}, \binits{S.K.}}:
\byear{2010},
\batitle{{Can Thermal Nonequilibrium Explain Coronal Loops?}}
\bjtitle{\apj}
\bvolume{714},
\bfpage{1239}.
\doiurl{10.1088/0004-637X/714/2/1239}.
\adsurl{2010ApJ...714.1239K}.
\end{barticle}
\endbibitem

\bibitem[\protect\citeauthoryear{{Klimchuk}, {Patsourakos}, and
  {Cargill}}{2008}]{Klimchuk2008}
\begin{barticle}
\bauthor{\bsnm{{Klimchuk}}, \binits{J.A.}},
\bauthor{\bsnm{{Patsourakos}}, \binits{S.}},
\bauthor{\bsnm{{Cargill}}, \binits{P.J.}}:
\byear{2008},
\batitle{{Highly Efficient Modeling of Dynamic Coronal Loops}}.
\bjtitle{\apj}
\bvolume{682},
\bfpage{1351}.
\doiurl{10.1086/589426}.
\adsurl{2008ApJ...682.1351K}.
\end{barticle}
\endbibitem

\bibitem[\protect\citeauthoryear{{Landi} \textit{et~al.}}{2013}]{Landi2013}
\begin{barticle}
\bauthor{\bsnm{{Landi}}, \binits{E.}},
\bauthor{\bsnm{{Young}}, \binits{P.R.}},
\bauthor{\bsnm{{Dere}}, \binits{K.P.}},
\bauthor{\bsnm{{Del Zanna}}, \binits{G.}},
\bauthor{\bsnm{{Mason}}, \binits{H.E.}}:
\byear{2013},
\batitle{{CHIANTI--An Atomic Database for Emission Lines. XIII. Soft X-Ray
  Improvements and Other Changes}}.
\bjtitle{\apj}
\bvolume{763},
\bfpage{86}.
\doiurl{10.1088/0004-637X/763/2/86}.
\adsurl{2013ApJ...763...86L}.
\end{barticle}
\endbibitem

\bibitem[\protect\citeauthoryear{{Lemen} \textit{et~al.}}{2012}]{lemen_2012}
\begin{barticle}
\bauthor{\bsnm{{Lemen}}, \binits{J.R.}},
\bauthor{\bsnm{{Title}}, \binits{A.M.}},
\bauthor{\bsnm{{Akin}}, \binits{D.J.}},
\bauthor{\bsnm{{Boerner}}, \binits{P.F.}},
\bauthor{\bsnm{{Chou}}, \binits{C.}},
\bauthor{\bsnm{{Drake}}, \binits{J.F.}},
\bauthor{\bsnm{{Duncan}}, \binits{D.W.}},
\bauthor{\bsnm{{Edwards}}, \binits{C.G.}},
\bauthor{\bsnm{{Friedlaender}}, \binits{F.M.}},
\bauthor{\bsnm{{Heyman}}, \binits{G.F.}},
\bauthor{\bsnm{{Hurlburt}}, \binits{N.E.}},
\bauthor{\bsnm{{Katz}}, \binits{N.L.}},
\bauthor{\bsnm{{Kushner}}, \binits{G.D.}},
\bauthor{\bsnm{{Levay}}, \binits{M.}},
\bauthor{\bsnm{{Lindgren}}, \binits{R.W.}},
\bauthor{\bsnm{{Mathur}}, \binits{D.P.}},
\bauthor{\bsnm{{McFeaters}}, \binits{E.L.}},
\bauthor{\bsnm{{Mitchell}}, \binits{S.}},
\bauthor{\bsnm{{Rehse}}, \binits{R.A.}},
\bauthor{\bsnm{{Schrijver}}, \binits{C.J.}},
\bauthor{\bsnm{{Springer}}, \binits{L.A.}},
\bauthor{\bsnm{{Stern}}, \binits{R.A.}},
\bauthor{\bsnm{{Tarbell}}, \binits{T.D.}},
\bauthor{\bsnm{{Wuelser}}, \binits{J.-P.}},
\bauthor{\bsnm{{Wolfson}}, \binits{C.J.}},
\bauthor{\bsnm{{Yanari}}, \binits{C.}},
\bauthor{\bsnm{{Bookbinder}}, \binits{J.A.}},
\bauthor{\bsnm{{Cheimets}}, \binits{P.N.}},
\bauthor{\bsnm{{Caldwell}}, \binits{D.}},
\bauthor{\bsnm{{Deluca}}, \binits{E.E.}},
\bauthor{\bsnm{{Gates}}, \binits{R.}},
\bauthor{\bsnm{{Golub}}, \binits{L.}},
\bauthor{\bsnm{{Park}}, \binits{S.}},
\bauthor{\bsnm{{Podgorski}}, \binits{W.A.}},
\bauthor{\bsnm{{Bush}}, \binits{R.I.}},
\bauthor{\bsnm{{Scherrer}}, \binits{P.H.}},
\bauthor{\bsnm{{Gummin}}, \binits{M.A.}},
\bauthor{\bsnm{{Smith}}, \binits{P.}},
\bauthor{\bsnm{{Auker}}, \binits{G.}},
\bauthor{\bsnm{{Jerram}}, \binits{P.}},
\bauthor{\bsnm{{Pool}}, \binits{P.}},
\bauthor{\bsnm{{Soufli}}, \binits{R.}},
\bauthor{\bsnm{{Windt}}, \binits{D.L.}},
\bauthor{\bsnm{{Beardsley}}, \binits{S.}},
\bauthor{\bsnm{{Clapp}}, \binits{M.}},
\bauthor{\bsnm{{Lang}}, \binits{J.}},
\bauthor{\bsnm{{Waltham}}, \binits{N.}}:
\byear{2012},
\batitle{{The Atmospheric Imaging Assembly (AIA) on the Solar Dynamics
  Observatory (SDO)}}.
\bjtitle{\solphys}
\bvolume{275}(\bissue{1-2}),
\bfpage{17}.
\doiurl{10.1007/s11207-011-9776-8}.
\adsurl{https://ui.adsabs.harvard.edu/abs/2012SoPh..275...17L}.
\end{barticle}
\endbibitem

\bibitem[\protect\citeauthoryear{{Lionello}
  \textit{et~al.}}{2013}]{Lionello2013}
\begin{barticle}
\bauthor{\bsnm{{Lionello}}, \binits{R.}},
\bauthor{\bsnm{{Winebarger}}, \binits{A.R.}},
\bauthor{\bsnm{{Mok}}, \binits{Y.}},
\bauthor{\bsnm{{Linker}}, \binits{J.A.}},
\bauthor{\bsnm{{Miki{\'c}}}, \binits{Z.}}:
\byear{2013},
\batitle{{Thermal Non-equilibrium Revisited: A Heating Model for Coronal
  Loops}}.
\bjtitle{\apj}
\bvolume{773},
\bfpage{134}.
\doiurl{10.1088/0004-637X/773/2/134}.
\adsurl{2013ApJ...773..134L}.
\end{barticle}
\endbibitem

\bibitem[\protect\citeauthoryear{{L{\'o}pez Fuentes} and
  {Klimchuk}}{2015}]{lopezfuentes_2015}
\begin{barticle}
\bauthor{\bsnm{{L{\'o}pez Fuentes}}, \binits{M.}},
\bauthor{\bsnm{{Klimchuk}}, \binits{J.A.}}:
\byear{2015},
\batitle{{Two-dimensional Cellular Automaton Model for the Evolution of Active
  Region Coronal Plasmas}}.
\bjtitle{\apj}
\bvolume{799}(\bissue{2}),
\bfpage{128}.
\doiurl{10.1088/0004-637X/799/2/128}.
\adsurl{https://ui.adsabs.harvard.edu/abs/2015ApJ...799..128L}.
\end{barticle}
\endbibitem

\bibitem[\protect\citeauthoryear{{L{\'o}pez Fuentes}, {Klimchuk}, and
  {D{\'e}moulin}}{2006}]{LopezFuentes2006}
\begin{barticle}
\bauthor{\bsnm{{L{\'o}pez Fuentes}}, \binits{M.C.}},
\bauthor{\bsnm{{Klimchuk}}, \binits{J.A.}},
\bauthor{\bsnm{{D{\'e}moulin}}, \binits{P.}}:
\byear{2006},
\batitle{{The Magnetic Structure of Coronal Loops Observed by TRACE}}.
\bjtitle{\apj}
\bvolume{639},
\bfpage{459}.
\doiurl{10.1086/499155}.
\adsurl{2006ApJ...639..459L}.
\end{barticle}
\endbibitem

\bibitem[\protect\citeauthoryear{{L{\'o}pez Fuentes}, {Klimchuk}, and
  {Mandrini}}{2007}]{LopezFuentes2007}
\begin{barticle}
\bauthor{\bsnm{{L{\'o}pez Fuentes}}, \binits{M.C.}},
\bauthor{\bsnm{{Klimchuk}}, \binits{J.A.}},
\bauthor{\bsnm{{Mandrini}}, \binits{C.H.}}:
\byear{2007},
\batitle{{The Temporal Evolution of Coronal Loops Observed by GOES SXI}}.
\bjtitle{\apj}
\bvolume{657},
\bfpage{1127}.
\doiurl{10.1086/510662}.
\adsurl{2007ApJ...657.1127L}.
\end{barticle}
\endbibitem

\bibitem[\protect\citeauthoryear{{Mandrini} \textit{et~al.}}{1996}]{Mandrini96}
\begin{barticle}
\bauthor{\bsnm{{Mandrini}}, \binits{C.H.}},
\bauthor{\bsnm{{D{\'e}moulin}}, \binits{P.}},
\bauthor{\bsnm{{van Driel-Gesztelyi}}, \binits{L.}},
\bauthor{\bsnm{{Schmieder}}, \binits{B.}},
\bauthor{\bsnm{{Cauzzi}}, \binits{G.}},
\bauthor{\bsnm{{Hofmann}}, \binits{A.}}:
\byear{1996},
\batitle{{3D Magnetic Reconnection at an X-Ray Bright Point}}.
\bjtitle{\solphys}
\bvolume{168}(\bissue{1}),
\bfpage{115}.
\doiurl{10.1007/BF00145829}.
\adsurl{https://ui.adsabs.harvard.edu/abs/1996SoPh..168..115M}.
\end{barticle}
\endbibitem

\bibitem[\protect\citeauthoryear{{Mandrini}
  \textit{et~al.}}{2014}]{Mandrini2014}
\begin{barticle}
\bauthor{\bsnm{{Mandrini}}, \binits{C.H.}},
\bauthor{\bsnm{{Schmieder}}, \binits{B.}},
\bauthor{\bsnm{{D{\'e}moulin}}, \binits{P.}},
\bauthor{\bsnm{{Guo}}, \binits{Y.}},
\bauthor{\bsnm{{Cristiani}}, \binits{G.D.}}:
\byear{2014},
\batitle{{Topological Analysis of Emerging Bipole Clusters Producing Violent
  Solar Events}}.
\bjtitle{\solphys}
\bvolume{289}(\bissue{6}),
\bfpage{2041}.
\doiurl{10.1007/s11207-013-0458-6}.
\adsurl{https://ui.adsabs.harvard.edu/abs/2014SoPh..289.2041M}.
\end{barticle}
\endbibitem

\bibitem[\protect\citeauthoryear{{Nuevo} \textit{et~al.}}{2015}]{Nuevo2015}
\begin{barticle}
\bauthor{\bsnm{{Nuevo}}, \binits{F.A.}},
\bauthor{\bsnm{{V{\'a}squez}}, \binits{A.M.}},
\bauthor{\bsnm{{Landi}}, \binits{E.}},
\bauthor{\bsnm{{Frazin}}, \binits{R.}}:
\byear{2015},
\batitle{{Multimodal Differential Emission Measure in the Solar Corona}}.
\bjtitle{\apj}
\bvolume{811},
\bfpage{128}.
\doiurl{10.1088/0004-637X/811/2/128}.
\adsurl{2015ApJ...811..128N}.
\end{barticle}
\endbibitem

\bibitem[\protect\citeauthoryear{{Parker}}{1988}]{parker_1988}
\begin{barticle}
\bauthor{\bsnm{{Parker}}, \binits{E.N.}}:
\byear{1988},
\batitle{{Nanoflares and the Solar X-Ray Corona}}.
\bjtitle{\apj}
\bvolume{330},
\bfpage{474}.
\doiurl{10.1086/166485}.
\adsurl{https://ui.adsabs.harvard.edu/abs/1988ApJ...330..474P}.
\end{barticle}
\endbibitem

\bibitem[\protect\citeauthoryear{{Pesnell}, {Thompson}, and
  {Chamberlin}}{2012}]{pesnell_2012}
\begin{barticle}
\bauthor{\bsnm{{Pesnell}}, \binits{W.D.}},
\bauthor{\bsnm{{Thompson}}, \binits{B.J.}},
\bauthor{\bsnm{{Chamberlin}}, \binits{P.C.}}:
\byear{2012},
\batitle{{The Solar Dynamics Observatory (SDO)}}.
\bjtitle{\solphys}
\bvolume{275}(\bissue{1-2}),
\bfpage{3}.
\doiurl{10.1007/s11207-011-9841-3}.
\adsurl{https://ui.adsabs.harvard.edu/abs/2012SoPh..275....3P}.
\end{barticle}
\endbibitem

\bibitem[\protect\citeauthoryear{{Reale}}{2014}]{Reale2014}
\begin{barticle}
\bauthor{\bsnm{{Reale}}, \binits{F.}}:
\byear{2014},
\batitle{{Coronal Loops: Observations and Modeling of Confined Plasma}}.
\bjtitle{Living Reviews in Solar Physics}
\bvolume{11},
\bfpage{4}.
\doiurl{10.12942/lrsp-2014-4}.
\adsurl{2014LRSP...11....4R}.
\end{barticle}
\endbibitem

\bibitem[\protect\citeauthoryear{{Rosner}, {Tucker}, and
  {Vaiana}}{1978}]{Rosner1978}
\begin{barticle}
\bauthor{\bsnm{{Rosner}}, \binits{R.}},
\bauthor{\bsnm{{Tucker}}, \binits{W.H.}},
\bauthor{\bsnm{{Vaiana}}, \binits{G.S.}}:
\byear{1978},
\batitle{{Dynamics of the quiescent solar corona}}.
\bjtitle{\apj}
\bvolume{220},
\bfpage{643}.
\doiurl{10.1086/155949}.
\adsurl{1978ApJ...220..643R}.
\end{barticle}
\endbibitem

\bibitem[\protect\citeauthoryear{{Scherrer}
  \textit{et~al.}}{2012}]{scherrer_2012}
\begin{barticle}
\bauthor{\bsnm{{Scherrer}}, \binits{P.H.}},
\bauthor{\bsnm{{Schou}}, \binits{J.}},
\bauthor{\bsnm{{Bush}}, \binits{R.I.}},
\bauthor{\bsnm{{Kosovichev}}, \binits{A.G.}},
\bauthor{\bsnm{{Bogart}}, \binits{R.S.}},
\bauthor{\bsnm{{Hoeksema}}, \binits{J.T.}},
\bauthor{\bsnm{{\etal}}}:
\byear{2012},
\batitle{{The Helioseismic and Magnetic Imager (HMI) Investigation for the
  Solar Dynamics Observatory (SDO)}}.
\bjtitle{\solphys}
\bvolume{275},
\bfpage{207}.
\doiurl{10.1007/s11207-011-9834-2}.
\adsurl{2012SoPh..275..207S}.
\end{barticle}
\endbibitem

\bibitem[\protect\citeauthoryear{{Schmelz}, {Christian}, and
  {Chastain}}{2016}]{Schmelz2016}
\begin{barticle}
\bauthor{\bsnm{{Schmelz}}, \binits{J.T.}},
\bauthor{\bsnm{{Christian}}, \binits{G.M.}},
\bauthor{\bsnm{{Chastain}}, \binits{R.A.}}:
\byear{2016},
\batitle{{The Coronal Loop Inventory Project: Expanded Analysis and Results}}.
\bjtitle{\apj}
\bvolume{831}(\bissue{2}),
\bfpage{199}.
\doiurl{10.3847/0004-637X/831/2/199}.
\adsurl{https://ui.adsabs.harvard.edu/abs/2016ApJ...831..199S}.
\end{barticle}
\endbibitem

\bibitem[\protect\citeauthoryear{{Schmelz} \textit{et~al.}}{2012}]{Schmelz2012}
\begin{barticle}
\bauthor{\bsnm{{Schmelz}}, \binits{J.T.}},
\bauthor{\bsnm{{Reames}}, \binits{D.V.}},
\bauthor{\bsnm{{von Steiger}}, \binits{R.}},
\bauthor{\bsnm{{Basu}}, \binits{S.}}:
\byear{2012},
\batitle{{Composition of the Solar Corona, Solar Wind, and Solar Energetic
  Particles}}.
\bjtitle{\apj}
\bvolume{755},
\bfpage{33}.
\doiurl{10.1088/0004-637X/755/1/33}.
\adsurl{2012ApJ...755...33S}.
\end{barticle}
\endbibitem

\bibitem[\protect\citeauthoryear{{Schmelz} \textit{et~al.}}{2015}]{Schmelz2015}
\begin{barticle}
\bauthor{\bsnm{{Schmelz}}, \binits{J.T.}},
\bauthor{\bsnm{{Pathak}}, \binits{S.}},
\bauthor{\bsnm{{Christian}}, \binits{G.M.}},
\bauthor{\bsnm{{Dhaliwal}}, \binits{R.S.S.}},
\bauthor{\bsnm{{Paul}}, \binits{K.S.}}:
\byear{2015},
\batitle{{The Coronal Loop Inventory Project}}.
\bjtitle{\apj}
\bvolume{813}(\bissue{1}),
\bfpage{71}.
\doiurl{10.1088/0004-637X/813/1/71}.
\adsurl{https://ui.adsabs.harvard.edu/abs/2015ApJ...813...71S}.
\end{barticle}
\endbibitem

\bibitem[\protect\citeauthoryear{{Winebarger}, {Warren}, and
  {Mariska}}{2003}]{Winebarger2003}
\begin{barticle}
\bauthor{\bsnm{{Winebarger}}, \binits{A.R.}},
\bauthor{\bsnm{{Warren}}, \binits{H.P.}},
\bauthor{\bsnm{{Mariska}}, \binits{J.T.}}:
\byear{2003},
\batitle{{Transition Region and Coronal Explorer and Soft X-Ray Telescope
  Active Region Loop Observations: Comparisons with Static Solutions of the
  Hydrodynamic Equations}}.
\bjtitle{\apj}
\bvolume{587},
\bfpage{439}.
\doiurl{10.1086/368017}.
\adsurl{2003ApJ...587..439W}.
\end{barticle}
\endbibitem

\end{thebibliography}

\end{article} 

\end{document}